\documentclass[acmlarge]{acmart}

\usepackage{enumitem}
\usepackage{appendix}
\usepackage{colortbl}
\usepackage{subcaption} 
\usepackage{fancybox}
\usepackage{multirow}
\usepackage{makecell}

\AtBeginDocument{%
  }

\setcopyright{cc}
\setcctype{by}
\acmJournal{IMWUT}
\acmYear{2025} \acmVolume{9} \acmNumber{3} \acmArticle{131} \acmMonth{9} \acmPrice{}\acmDOI{10.1145/3749519}

\begin{document}


\title{AnnoSense: A Framework for Physiological Emotion Data Collection in Everyday Settings for AI}


\author{Pragya Singh}
\orcid{0000-0003-3933-2224}
\affiliation{%
  \institution{IIIT-Delhi}
  \city{New Delhi}
  \country{India}}
\email{pragyas@iiitd.ac.in}

\author{Ankush Gupta}
\orcid{0009-0006-0596-5436}
\affiliation{%
\institution{IIIT-Delhi}
  \city{New Delhi}
  \country{India}}
\email{ankush21232@iiitd.ac.in }

\author{Mohan Kumar}
\orcid{0000-0002-0286-6997}
\affiliation{%
\institution{RIT}
  \city{Rochester}
  \country{New York, US}}
\email{mjkvcs@rit.edu}

\author{Pushpendra Singh}
\orcid{0000-0003-2152-1027}
\affiliation{%
\institution{IIIT-Delhi}
  \city{New Delhi}
  \country{India}}
\email{psingh@iiitd.ac.in}

\begin{abstract}
Emotional and mental well-being are vital components of quality of life, and with the rise of smart devices like smartphones, wearables, and artificial intelligence (AI), new opportunities for monitoring emotions in everyday settings have emerged. However, for AI algorithms to be effective, they require high-quality data and accurate annotations. As the focus shifts towards collecting emotion data in real-world environments to capture more authentic emotional experiences, the process of gathering emotion annotations has become increasingly complex. This work explores the challenges of everyday emotion data collection from the perspectives of key stakeholders. We collected 75 survey responses, performed 32 interviews with the public, and 3 focus group discussions (FGDs) with 12 mental health professionals. The insights gained from a total of 119 stakeholders informed the development of our framework, \textit{AnnoSense}, designed to support everyday emotion data collection for AI. This framework was then evaluated by 25 emotion AI experts for its clarity, usefulness, and adaptability. Lastly, we discuss the potential next steps and implications of \textit{AnnoSense} for future research in emotion AI, highlighting its potential to enhance the collection and analysis of emotion data in real-world contexts.
\end{abstract}




\maketitle

\section{Introduction}

In today’s fast-paced world, maintaining good mental health has become increasingly important. The challenges associated with monitoring emotional burnout and stress often lead to significant mental health issues and a diminished quality of life. Recent advancements in ubiquitous computing, wearable devices and mobile phones equipped with sensors for tracking physiological or behavioral changes and artificial intelligence (AI) algorithms are creating new opportunities for monitoring mental well-being \cite{xu_globem_2022, wang2014studentlife}. Several wearable and mobile phone-based interventions have been designed to monitor stress, sleep, mood, habits, and emotions \cite{hickey2021smart, perez2021wearables, traunmuller2024wearablehealthcaredevicesmonitoring}. These bio-signal data-driven technologies have the potential to support continuous monitoring, providing capabilities for early diagnosis and enabling data-driven insights for mental health professionals \cite{tsirmpas2022feasibility}. 
Recent commercial developments highlight the increasing integration of physiological sensors into mainstream wearable devices. Consumer products such as smartwatches and smart rings now commonly include sensors for Photoplethysmography (PPG), Electrodermal Activity (EDA), and Skin Temperature (SKT), in addition to standard accelerometer and gyroscope components. These combined features enable more comprehensive emotion and stress assessments. For example, the Oura Ring categorizes stress into four states—stressed, engaged, relaxed, and restored—to help users monitor, understand, and manage daily stressors \cite{neigel2024using, ong2021longitudinal, antikainen2024acute, nagaraj2023dissecting}. Major wearable brands such as Apple \cite{applewatch2024}, Samsung \cite{samsungwatch2024}, Fitbit \cite{fitbit2024}, and Garmin \cite{garmin2024} have incorporated stress and mental well-being tracking into their devices and companion apps. Products like the Oura Ring \cite{ouraring2024}, Apple Watch \cite{applewatch2024}, WHOOP \cite{whoop2024}, and Samsung wearables \cite{samsungwatch2024} now also feature emotion journaling, reflecting a growing emphasis on physiological approaches to mental health monitoring. Moreover a new category of devices physiological-sensing devices called \textit{Earables} are also emerging \cite{10.1145/3550314}.
In addition to wearables, mobile phone applications are playing an increasingly important role in tracking emotions and mental well-being. Notable examples include Wysa \cite{wysa2024}, Woebot \cite{woebot2024}, Calm \cite{calm2024}, Daylio \cite{daylio2024}, and Headspace \cite{headspace2024}, which provide tools for emotion monitoring, cognitive support, and mindfulness.

Physiological signal based-emotion tracking still remains in its early stages, especially when it comes to continuous and reliable monitoring in everyday life \cite{hickey2021smart}. While there is growing commercial interest and increasing integration of physiological sensors in consumer wearables, current systems are limited to stress tracking and are lacking the sensitivity, personalization, and contextual awareness needed to fully capture the complexity of human emotions in real-world settings. AI-powered solutions that combine wearable devices with mobile phone applications present a promising approach to addressing these challenges, especially given recent advancements and the growing adoption of AI in tackling complex, real-world problems. However, a major limitation in developing such models is their heavy reliance on high-quality, labeled emotion data that accurately reflects the nuances of human emotional experiences \cite{9466092, 9779458, 10.1145/3711093, 6553804}. Most existing emotion datasets are collected in controlled laboratory environments or through short-term studies using self-report methods or expert-annotations. These approaches frequently fall short of capturing the dynamic, context-dependent, and multi-layered nature of emotions as they naturally unfold \cite{gao2023critiquing, 10.1145/3711093, das2022semantic, 10.1145/3491102.3517453, singh2024saycatcatunderstanding}. Consequently, models trained on such data often struggle to generalize to real-world scenarios, limiting their effectiveness and reliability for end users \cite{gao2023critiquing, 9779458, 10.1145/3341162.3349573, 9466092}, suggesting a need for methods to accurately capture emotional annotations and contextual information in real-life settings. Prior work in real-life settings has primarily relied on self-assessment approaches, such as the Experience Sampling Method (ESM)—which prompts users to report their emotions at regular or random intervals throughout the day—and the Day Reconstruction Method (DRM), where participants provide retrospective reports of their emotional states via structured questionnaires at the end of the day \cite{10.1145/3334480.3383019, 10.1016/j.ijhcs.2018.12.002, kosch2020emotions, schneider2020comparability, stone2006population}.
While useful for capturing momentary or reflective self-reports, these methods often provide labels using predefined emotional scales (such as, Self-Assessment Manikin (SAM) \cite{bradley1994measuring} or scales based on six basic emotions \cite{ekman1992there}), offering limited insights on contextual information about emotional experiences. 
Additionally, previous studies have highlighted further challenges; for instance, the act of annotation itself may influence the user’s emotional state, introducing measurement bias \cite{10.1145/3491102.3501944}. Moreover, participants often experience annotation fatigue, leading to low motivation and engagement over time \cite{10.1145/3411764.3445771}. Furthermore, prior research has also emphasized the need for ecologically valid, human-centric data collection approaches that reflect how emotions are naturally experienced in everyday contexts \cite{9779458, 10.1145/3711093, gao2023critiquing} These insights suggests a significant gap in studies that investigate in existing methodologies from participant perspectives, who are not only the sources of data but also its annotators \cite{10.1145/3711093, gao2023critiquing}. 

To explore the challenges and identify opportunities in everyday emotion annotation methods for physiological signal-based emotion AI (referred to as "emotion AI" in this paper) research using wearables and mobile phone data, we designed this study. Our approach centers on examining the perspectives of diverse stakeholders, including both users and non-users of emotion-tracking technologies, as well as mental health professionals. Through this lens, we aim to understand the human side of emotion data collection and its implications for designing more effective and user-centric emotion AI systems. Specifically, we address the following research questions:

\textbf{RQ1:} What are participants’ perspectives on the challenges and opportunities for annotating (identifying and labeling) and tracking emotions in their daily lives?

\textbf{RQ2:} What are the perspectives of mental health professionals on the challenges and opportunities in physiological emotion data collection for developing emotion AI interventions?

\textbf{RQ3:} How can participants' and domain experts' perspectives be integrated to develop a holistic methodology for collecting physiological emotion data in real-life settings?


We employed a qualitative research method, including surveys (n = 75) and interviews (n = 32) with members of the public (with and without experience in therapy or counseling, as well as those who have used wearables and mobile phone applications for tracking stress and emotions, or participated in emotion data collection studies), as well as focus group discussions (n = 3) involving 12 mental health professionals. Following our methodology, our study evaluated the needs of participants from diverse perspectives, covering a total sample size of 119 participants. In this study, \textit{emotion} is defined in line with the "Theory of Constructed Emotion" as proposed by Lisa Barrett \cite{barrett2017emotions}, which views emotions - "as individualized, context-dependent experiences constructed by the brain through the interpretation of bodily sensations (e.g., heart rate, arousal) in relation to past experiences, situational context, and learned emotional concepts, not as fixed biological responses". In contrast with the prior research where \textit{emotions} are typically modeled as changes in physiological response and behavioral reaction, using physiological signals (e.g., heart rate, skin temperature, electrodermal activity), behavioral patterns, and self-reported data \cite{picard1997w}. By adopting Barrett's perspective, this study emphasizes the context-dependent and dynamic nature of emotions, enabling a more comprehensive exploration of emotional experiences.
Also, for this study, we have referred to structured methods, such as scales like PANAS, SAM, or Likert scales, as objective methods, while unstructured methods, such as providing an option to write, audio record, or add images, are referred to as subjective methods. This human-centric view of \textit{emotions} helps us to study the subjectivity and variability of emotional experiences, challenging the assumption that specific physiological or behavioral patterns can map directly to discrete or dimensional emotion categories \cite{singh2024saycatcatunderstanding}. Our paper makes the following key contributions:

\begin{itemize}
    \item \textbf{Annosense Framework:} We introduce Annosense, a novel framework comprising 15 actionable guidelines for collecting well-annotated wearable and mobile-based emotion data in everyday settings. These guidelines are derived from an in-depth analysis of user experiences, contextual challenges, and common challenges in emotion data collection.
    \item \textbf{Expert Evaluation}: We evaluate the Annosense framework through feedback from 25 emotion AI experts with professional and academic experience, making this the first work to present evaluated guidelines tailored specifically for real-world emotion data collection.
    \item \textbf{Potential Implementation}: We identify next steps for designing participant-aware systems in practice, informed by the Annosense framework and a review of current tools, technologies, and applications.
    \item \textbf{Design Implications}: Through our findings and expert discussions, we offer design recommendations for future emotion data collection practices and AI algorithm development.
\end{itemize}

Finally, it is crucial to note that this work introduces a new paradigm of designing emotion data-collection studies from participants' and domain experts' perspectives. These contributions are significant for the Ubiquitous Computing (UbiComp), Human Computer Interaction (HCI), and affective computing communities as they address a critical gap in how emotional data is collected and validated in real-world, everyday settings, moving beyond controlled lab environments. By offering a systematically developed and expert-evaluated framework, this work equips researchers and practitioners with practical guidelines that are grounded in user context, data-centric AI practices, and ethical considerations. This not only advances methodological rigor in affective computing but also aligns with HCI’s emphasis on human-centered design, participatory development, and context-sensitive technologies. Moreover, by outlining pathways for implementation, the work bridges theory and practice, supporting the development of emotion-aware technologies that are both technically sound and socially responsible.



\section{Related Work} 

\subsection{Understanding Emotions - From Emotion Theories to Frameworks} 

The definition of Emotions has been a widely debated topic among researchers across various domains, including philosophy, psychology, and neuroscience. It has evolved over time alongside the development of different theoretical perspectives on what emotions are and how they function \cite{tian2022applied}. Early theories of emotions have defined emotions as an outcome of evolution \cite{plutchik1982psychoevolutionary}, where theorists have classified emotions into distinct sets of basic emotions that are universal. In Ekman's Basic Emotion Theory \cite{ekman1992there}, emotions have been classified into six basic emotions - anger, fear, disgust, happiness, surprise, and sadness - that are biologically hardwired and are universally recognized. In contrast, Appraisal theories, pioneered by psychologists like Richard Lazarus \cite{lazarus1991progress}, defined emotions as the result of individual cognitive appraisals of events, accounting for subjectivity in emotional responses, and suggesting emotions as a response to how a person interprets a situation. More recently, Lisa Feldman Barrett's "Theory of Constructed Emotion" \cite{barrett2017emotions} proposed emotions as an active construction by the brain based on a combination of sensory input, past experiences, and cultural learning, and not something universal. Within computational approaches, emotions are often defined as complex psychological states with key components such as subjective experiences and behavioral and physiological responses. 
This lack of theoretical consensus has significant implications for physiological signal-based emotion AI. Unlike fields such as computer vision and NLP, where data modalities (e.g., images, videos, text) are well-defined, and labeling methods are standardized as per the task, such as classification or segmentation. Data in Emotion AI vary widely both in terms of the signal modalities collected, ranging from ECG, PPG, EDA, and heart rate variability to EEG, EMG, and fMRI, and the frameworks applied for emotion annotation (e.g., discrete categories, dimensional models like valence-arousal, hybrid or custom approaches). As a result, generalization across models is limited, reproducibility suffers, and the lack of interoperability between datasets complicates benchmarking and progress \cite{10.1145/3442188.3445939, 8217969, 10.1145/3432220, prajod2024stressortypematters}.

Despite these challenges, there are growing efforts to impose structure through methodological guidelines, annotation standards, and sensing toolkits. Researchers in computer science have borrowed standardized self-reporting scales and questionnaires from psychology for emotion annotations. Some of the most widely used scales include the Positive and Negative Affect Schedule (PANAS), which asks participants to rate how they experience different emotions \cite{watson1988development}, and the Self-Assessment Manikin (SAM), which helps users to rate their feelings along dimensions like valence (pleasant–unpleasant), arousal (excited–calm), and dominance \cite{bradley1994measuring}. 
These tools provide a structured way for researchers to capture emotional states. Further, researchers have also developed frameworks for keeping the interdisciplinary nature of emotion research in mind. For instance, frameworks like the HUMAINE project \cite{douglas2007humaine} and Emotion Annotation and Representation Language (EARL) \cite{schroder2006first} were developed to offer a common format for labeling emotions across different modalities such as speech, facial expression, and physiological signals, helping standardize how emotions are defined and interpreted in computational systems. Additionally, these frameworks provide guidelines for collecting emotion annotations in different use cases. Beyond these initial frameworks, for real-world emotion data collection, researchers have developed digital phenotyping and mobile sensing tools for context-aware ecological momentary assessments (EMAs). Popular tools includes, MindLamp \cite{vaidyam2022enabling}, Beiwe \cite{onnela2021beiwe}, AWARE \cite{ferreira2015aware}, AWARE-Light \cite{van2023aware}, PACO \cite{paco}, Sensingkit \cite{katevas2016sensingkit}, mEMA \cite{ilumivu_mema}, Experiencesampler \cite{thai2018experiencesampler}, and MobileQ \cite{meers2020mobileq}. These tools provide a shared structure for sensor integration, self-reports, and temporal alignment of signal data. However, prior research has still emphasized the challenge in grounding emotion data for machine learning algorithms, pointing out the need for adding participants' context to emotion data \cite{10.1145/3491102.3517453, 10.1145/3711093}. Further researchers have also cautioned about the biases that emotion data contains and its impact on data quality \cite{das2022semantic}. Moreover, the need to enhance the methodological aspects of emotion data collection was also highlighted for improving the overall data quality \cite{gao2023critiquing}. This points to the significance of further improving on the nuances of current approaches by integrating experts' opinions and participants ' specific factors into the data collection methodology. However, an in-depth exploration of how both participants and domain experts perceive these methods remains lacking, highlighting the need for further work in bridging the gap between structured approaches and meaningful participant engagement.


\begin{table*}[ht]
\centering
\begin{tabular}{p{3.5cm}p{4.5cm}p{4.5cm}}
\hline
\textbf{Dataset} & \textbf{Elicitation Method} & \textbf{Annotation Approach} \\
\hline
\textbf{WESAD} \cite{10.1145/3242969.3242985} & Video Clips, Public speaking, mental arithmetic, and Meditation & PANAS, SAM,  State-Trait Anxiety Inventory (STAI), Short Stress State Questionnaire
(SSSQ), physiological signals \\
\hline
\textbf{ASCERTAIN} \cite{10.1109/TAFFC.2016.2625250} & Video Clips & Valence-Arousal, Engagement, Liking, Familiarity, Personality Traits\\
\hline
\textbf{CASE} \cite{10.1038/s41597-019-0209-0} & Video Clips & Continuous Valence-Arousal Annotations\\
\hline
\textbf{Neurological Status} \cite{10.1109/SiPS.2016.27} & Physical Activities & Task-based Labels \\
\hline
\textbf{CLAS} \cite{10.1109/BIA48344.2019.8967457} & Video Clips, Images, Math, Stroop, Logic tasks & Arousal-Valence, Task-based Labels\\
\hline
\textbf{VREED} \cite{tabbaa2021vreed} & VR Video Clips & SAM, PANAS \\
\hline
\textbf{POPANE} \cite{10.1038/s41597-021-01117-0} & Speech preparation, Anticipation task, Interpersonal communication, Affective Images, and Video Clips & Discrete Emotion Categories, SAM, Avoidance Approach Motivation \\
\hline
\textbf{EMOGNITION} \cite{10.1038/s41597-022-01262-0} & Audio-visual stimuli & Discrete Emotion Categories, SAM, Avoidance Approach Motivation \\
\hline
\textbf{StressID} \cite{NEURIPS2023_5f09bfe6} & Cognitive load tasks & SAM, Custom Perceived Stress Assessment \\
\hline
\textbf{BIRAFFE2} \cite{10.1038/s41597-022-01402-6} & Games, Affective Music, and Images & SAM, Game Experience Questionnaire (GEQ) \\
\hline
\textbf{EEVR} \cite{singh2024eevr} & VR Video Clips & Textual Descriptions, SAM, PANAS, Familiarity, Liking, Personality Traits\\
\hline
\textbf{KEMOCON} \cite{park2020k} & 10-minute-long debate on social issues & Self-report Valence-Arousal, Discreet Emotion Category, Partner Annotations, and Expert Annotation \\
\hline
\textbf{AMIGOS} \cite{miranda-correa_amigos_2021} & Video clips (Long and Short) & SAM, PANAS, Personality traits \\
\hline
\textbf{RECOLA} \cite{ringeval_introducing_2013} & Collaborative task (video chat) & SAM, PANAS \\
\hline
\end{tabular}
\caption{Lab-based emotion datasets: Elicitation methods, annotation strategies, and labeling approaches}
\label{tab:lab_datasets}
\end{table*}

\subsection{Emotion Data Practices}\label{datacollectionrw}

Prior research in emotion and affect recognition has utilized a range of computational methods to infer emotional states, often relying on various data proxies \cite{picard1997w}. Within the fields of ubiquitous computing and artificial intelligence, a variety of proxy data modalities have been explored to infer emotions, including physiological signals (e.g., heart rate, electrodermal activity), facial expressions, vocal characteristics, mobile sensor data, and textual data \cite{10.1109/ACCESS.2021.3068045, 10.3390/s21165554}. This study specifically focuses on physiological signals and mobile sensing data, as these modalities offer a valuable combination of being continuously collectible and non-invasive. Their unobtrusive nature makes them particularly suitable for long-term, real-world emotion monitoring, enabling them to capture emotions with minimal participant disruption. In this section, we will discuss in detail how emotion data is collected and annotated in various settings. 

\textbf{Lab-Based Emotion Datasets:} The primary reason for collecting emotion data in lab settings is the amount of control it provides over stimulus presentation and participant conditions \cite{10.1109/TAFFC.2016.2625250, 10.1145/3242969.3242985, 10.1038/s41597-019-0209-0}. As shown in Table \ref{tab:lab_datasets}, a variety of emotion elicitation techniques and annotation approaches have been explored by researchers within lab settings, including elicitation methods like images, videos, audio, virtual reality (VR), cognitive tasks, and physical activities. While these datasets provide rich multimodal records of emotions, their methodological choices vary widely. Moreover, many of these methods rely on rigid self-reporting methods such as PANAS \cite{watson1994panas}, SAM \cite{bradley1994measuring}, or scales/emotion categories based on evolutionary theories of emotions like Ekman's Basic Emotions \cite{ekman1992there}, Kazemzadeh's 20 categories \cite{kazemzadeh2011emotion}). Further, there is minimal opportunity for participants to provide context on their emotional states \cite{10.1145/3711093}.

\textbf{Real-life and Constrained Emotion Datasets:} Emotion data collection in real-life and constrained settings offers a balance between ecological validity and experimental control, contributing to a deeper understanding of emotional states in diverse contexts as shown in Table \ref{tab:realdatset}. Constrained task-based studies such as ForDigitStress \cite{2303.07742}, NURSE \cite{10.1038/s41597-022-01361-y}, and G-REx \cite{bota2024real} often involve structured environments like job interviews, healthcare shifts during COVID-19, or prolonged exposure to emotion-eliciting media, where emotions are annotated through self-reports, physiological markers, or external ratings. These studies benefit from higher control over task and timing but face challenges such as context-driven label bias (like, Nurse dataset contains mostly negative emotion data due to the collection setting) and limited participant context in the annotation. On the other hand, semi-naturalistic datasets collected within constraint environments like colleges or workplaces such as StudentLife \cite{wang2014studentlife}, Laureate \cite{10.1145/3610892}, GLOBEM \cite{xu_globem_2022}, DiversityOne \cite{10.1145/3712289}, TILES \cite{yau2022tiles}, and the SWEET Study \cite{10.1038/s41746-018-0074-9} rely on Ecological Momentary Assessment (EMA) for self-reports alongside other survey information and sensor data to capture context. Further contributions, such as DAPPER \cite{10.1038/s41597-021-00945-4} and K-EmoPhone \cite{kang2023k}, explore daily-life emotional states through intensive prompting or periodic logging. 

\begin{table*}[ht]
\centering
\begin{tabular}{p{3.5cm} p{4.5cm} p{4.5cm}}
\toprule
\textbf{Dataset} & \textbf{Context/Task} & \textbf{Annotation Method} \\
\midrule
\textbf{ForDigitStress} \cite{2303.07742} & Job interview tasks simulating time pressure & Custom Stress Scale and Saliva Cortisol \\
\hline
\textbf{NURSE} \cite{10.1038/s41597-022-01361-y} & Healthcare workers during COVID-19 & Custom Stress Questionnaire \\
\hline
\textbf{G-REx} \cite{bota2024real} & Long movie viewing sessions & Post-Hoc SAM Scale Based Tool \\
\hline
\textbf{Laureate} \cite{10.1145/3610892} & University setting with student academic routines & Custom EMA (PANAVA-KS, physical activity, breakfast ingestion, caffeine intake, study-time and sleep quality) \\
\hline
\textbf{StudentLife} \cite{wang2014studentlife} & University campus life over multiple weeks & Photographic Affect Meter (PAM) EMA, Single-item Stress EMA \\
\hline
\textbf{GLOBEM} \cite{xu_globem_2022} & Naturalistic daily experiences across diverse locations & EMA Survey (PHQ-4, PSS-4, PANAS), and Pre-Post Survey \\
\hline
\textbf{TILES} \cite{yau2022tiles, yau2022tiles} & Workplace monitoring in hospital environment & Single-item Stress EMA, Survey on daily stressors, work behaviors, and sleep \\
\hline
\textbf{DAPPER} \cite{10.1038/s41597-021-00945-4} & Daily life across varied settings (field study) & 20-Item ESM (Information about daily events, Participants' openness to sharing emotion, TIPI-C, PANAS), DRM with Open-ended Question \\
\hline
\textbf{K-EmoPhone} \cite{kang2023k} & Daily life across varied settings (field study) & Custom Questionnaire (Valence, Arousal, Attention, Stress, Emotion Duration, Task Disturbance, Emotion Change)\\
\hline
\textbf{SWEET Study} \cite{10.1038/s41746-018-0074-9} & Office workers’ daily routines in real-life settings &  EMA (Stress, Activity, Food and Beverage Consumption, Sleep Quality, and Gastro-intestinal Symptoms) \\
\bottomrule
\end{tabular}
\caption{Tasks and Annotation Methods in Semi-Controlled Emotion Datasets.}
\label{tab:realdatset}
\end{table*}

While these approaches yield high ecological realism, they also introduce challenges such as lower response rates and reactivity to prompts. Additionally, self-reports collected within natural settings often lack enough information for reliably contextualizing the collected self-reports \cite{10.1145/3460418.3479338, gao2023critiquing}. Despite methodological innovations, a key limitation across both types of data collection is the reliance on approaches designed to simply collect data without keeping participants' factors in mind \cite{10.1145/3711093, das2022semantic}, which often led to quality issues in datasets. Recently, methods based on appraisal theories and constructive theories \cite{viola2021constructivist} have also been explored for labeling emotion data to capture more context-dependent labels \cite{israel2019emotion, larradet2019appraisal}. However, it remains in an initial phase, suggesting the need for more research in designing data collection methods within everyday settings.



\subsection{Ubiquitous Interventions for Emotion Annotations and Self-Reporting}

Recently, ubiquitous computing and related communities have begun exploring interactive approaches for labeling emotional data \cite{8668435, rajcic2020mirror, wang2018mirroru, hook2009affective}. Notable works on emotional annotation include "Find the Bot" \cite{10.1145/3613904.3642880}, which uses a web-based gaming platform to collect emotion annotations for machine learning algorithms, Reconexp \cite{10.1145/1409240.1409316}, where participants were provided with a both mobile and web-based interface, mirrorU \cite{10.1145/3170427.3188517}, which supports reflective writing through memory-based cues. Similarly, several other emotion measurement tools have been developed to support self-reporting through structured formats. These include the Affect Grid \cite{russell1989affect}, the Differential Emotions Scale \cite{boyle1984reliability}, and interactive tools such as Premo \cite{desmet2003measuring} and the Photographic Affect Meter (PAM) \cite{10.1145/1978942.1979047}, where users select images that best represent their emotional state. Researchers have recently designed an interactive mobile version for the Geneva emotion wheel to support emotion self-reporting \cite{simonazzi2021geneva}.
Further, prior works like PResUP \cite{10.1145/3678569}, a framework that probes users to self-report emotions opportunistically, and Mirror Ritual \cite{10.1145/3313831.3376625}, which uses facial recognition to detect participants' emotions and generate poems to encourage emotional reflections, are also explored. Further techniques, like Diurnal Rhythms of Emotions, based on circadian rhythms \cite{stone2006population}, Technology-Assisted Reconstruction (TAR) \cite{karapanos2012beyond} where passively collected data is used in assisting the later annotations at the end of the day \cite{10.1038/s41597-021-00945-4}, and Mirror Hearts \cite{10.1145/3544549.3585607}, where AI-powered third-person view is leveraged for self-reporting emotions were also explored. 
These approaches aim to provide more accurate representations of human emotions, moving beyond the limitations of reductionist models and scale-based labels. Recently, LLM-based self-reporting and in-context journaling have also been explored for behavioral monitoring in everyday settings, including Dairyhelper \cite{li2024diaryhelperexploringuseautomatic}, Mindshift \cite{10.1145/3613904.3642790}, and Mindscape \cite{10.1145/3699761}.
Despite ongoing work, emotion annotations in everyday settings remain challenging, as most of these methods are not translated for data collection methodology in noisy real-life settings. This paper aims to address this translation gap from stakeholders' (users and mental health professionals) perspectives to design novel participant-centric methods for emotion data collection in everyday settings.

\subsection{Emotion Monitoring and Stakeholders' Perspective}

Emotions have been studied for decades to improve human-machine or human-human interactions. Emotion recognition to support mental well-being \cite{thieme2020machine, 10.1145/3544548.3581209}, employee well-being, productivity \cite{10.1145/3544548.3580950}, individual monitoring, behavior tracking, and emotion regulation \cite{slovak2023designing, bakker2018engagement, costa2019boostmeup} have been studied in the past. Prior works have also explored data collection practices within AI from multiple stakeholders' points of view \cite{10.1145/3411764.3445518}, highlighting the assumptions about data being something that is readily available to be used \cite{10.1145/3491101.3503724, 10.1145/3411764.3445518}. However, this attitude of considering data as something readily available has often led to poor data quality and degraded the performance of AI algorithms. In the past, mental health monitoring solutions were scrutinized from users' perspectives \cite{wani2024unrest, 10.1145/3025453.3025750, 10.1145/3411764.3445771}. Kelley et al., who work on students' mental health, highlighted the challenges of self-tracking \cite{10.1145/3025453.3025750, 10.1145/3411764.3445771} and found motivation to be a major challenge due to factors, such as fear towards tracking negative emotion data. Further, Zhang et al. \cite{10.1145/3411764.3445771} studied the experiences of users suffering from depression and anxiety in using mental health tracking applications and highlighted that the use of customization (such as creating checklists for daily progress) is correlated with symptom severity. Another common theme among prior works on users' attitudes was the doubt towards the authenticity of digital tools as compared to humans, which can provide real-world care and social support \cite{10.1145/3313831.3376362, 10.1145/3476049, 10.1145/3491102.3517498}. Further prior work on the perceived utility of wearables for mental well-being \cite{10.1145/3544548.3581209} highlighted users' attitudes towards tracking as per needs, for instance, tracking for maintenance, for people with few symptoms, versus tracking for active symptom management, for people with more symptoms. Studies have also highlighted the users' views on the benefits of using mental health applications, such as support in identifying patterns, better emotional awareness, and emotional regulation \cite{10.1145/3491102.3517498}. However, challenges remain around potential drawbacks or adverse effects that "misdiagnosis" and "failure or error in delivering important messages" can have on users \cite{10.1145/3491102.3517498}. Further emotion recognition on social media, work environments, and daily life emotional tracking \cite{10.1145/3313831.3376680, 10.1145/3491102.3517498, 10.1145/3544548.3580950, 10.1145/3579600, 10.1145/3491102.3517498} has also been perceived as invasive, frightening, and associated with a loss of autonomy or control, suggesting the importance of considering the sensitivity of emotion data. Moreover, Singh et al. \cite{10.1145/3711093}, Swain et al. \cite{das2022semantic}, and Gao et al. \cite{gao2023critiquing} have highlighted the influence that participants' self-reporting can have on the data quality. Prior work has also explored machine learning models for prompting users to annotate emotions as per their physiological data \cite{9431143}. However, the performance of the pre-trained model remained at a subpar level due to the unavailability of quality data that is representative of real-life settings.
The challenges around the usability of mental health tracking and data quality suggest the importance of careful maneuvering for emotion data collection methods.
It is crucial for several reasons, as annotations play an important role in overall data quality. In the case of emotion data, it remains further significant due to the absence of a global gold-standard definition of emotions \cite{10.1145/3442188.3445939, singh2024saycatcatunderstanding}. Thus, in this work, we aim to explore these gaps in-depth to get a holistic view of stakeholders' perspectives on emotion data collection for AI. 

\section{Methodology}
We employed a qualitative research approach, utilizing surveys, interviews, and focus group discussions to gather diverse perspectives from our stakeholders. 
This study received approval from the Institutional Review Board (IRB) at IIIT-Delhi, India, to ensure ethical compliance. Participation (Survey, Interview, FGD, and Guideline Evaluation) was entirely voluntary, and no compensation was offered to participants. The following subsections provide a detailed explanation of our employed methods.

\subsection{Survey}\label{survey_descrip}

To answer our research questions, we surveyed individuals aged 18 and above, including people with or without experience with emotion tracking technologies. Our survey was developed as an exploratory, mixed-methods tool to investigate how users perceive, interpret, and prefer to annotate emotional experiences in their daily lives. It was designed with reference to established emotion theories \cite{barrett2001knowing, barrett2017emotions} and user-centered design principles \cite{baxter2015understanding, lazar2017research}, including a mix of quantitative and open-ended questions to capture both structured responses and rich personal narratives (see appendix \ref{surveyapp} for detailed questionnaire). It comprised 27 questions organized into three main sections: \textbf{1) Demographic Details}: This section collected basic demographic information about our study participants, \textbf{2) Understanding Emotional Awareness}: This section was designed to explore how participants perceive, differentiate, and articulate their emotional experiences and was grounded in the concepts of emotional awareness, emotional vocabulary, and emotional granularity \cite{barrett2001knowing, barrett2017emotions}. To assess emotional awareness, participants were asked questions such as, “How often do you take time to reflect on your emotions?”, “When experiencing a strong emotion, how easily can you identify what emotion you are feeling?” and “How often do you feel mixed emotions?” Further, they were asked to reflect on which emotions they find easier or harder to identify and to list five positive and/or negative emotions they commonly experience, along with their impact on daily life. These responses helped us understand each participant’s emotional vocabulary and how they articulate emotional states. To further assess emotional granularity— the ability to distinguish between similar emotions— participants were asked whether they could tell emotions like sadness and disappointment or anger and frustration apart, and to explain their reasoning. This provided insight into their ability to make fine-grained emotional distinctions linked to more effective emotional regulation and self-awareness. In addition, participants were prompted to describe any recent situations involving strong emotions and to identify the emotions they experienced, to evaluate further their ability to identify and express emotions linguistically \cite{lindquist2014emotion}. We also included questions to assess the conceptual understanding of important terms like emotions and emotional intensity. To assess conceptual understanding, we included targeted items such as a multiple-choice question asking participants to define “emotion” (e.g., as a bodily sensation, mental state, or response to external events). Further, we also asked them to define “emotional intensity” in their own words. Further, participants were asked about their previous experience with emotion management and use of tools or real-life techniques, such as wearables, emotion-tracking applications, mindfulness practices, and journaling, that assist them in identifying, labeling, and regulating emotions.
\textbf{3) Attitudes Toward Daily Emotion Annotation}: This section examined how users feel about incorporating emotion annotation into their daily routines. It included questions assessing the willingness to annotate positive and negative emotions, preferred annotation methods, and perceived barriers. Example questions included: “Would you like to annotate your emotions daily?”, “What factors are most important to consider when labeling emotions?”, and “How easy do you find it to annotate your emotions daily?” Participants were also asked to express their annotation preferences (e.g., emoji, voice input, or descriptive text) and their preferred frequency to annotate emotions daily.

To maintain the quality of our survey responses, we included several consistency checks in our questionnaire, where users were prompted to explain their choices qualitatively for multiple-choice and Likert-type questions. Further, our survey contained 14 open-ended questions, which also enhanced the depth and authenticity of our survey data. Additionally, to validate our survey design for question clarity, logical flow, and completion time, we conducted a pilot with 6 participants before our data collection. Moreover, to evaluate the quality of the responses to our open-ended survey questions, we calculated completion rates to assess participant engagement, distinguishing between required and optional questions. Additionally, we examined word count statistics for each open-ended question, including range, mean, and standard deviation, as proxies for response depth and variation (see Table \ref{tab:survey-quality}). There were 14 open-ended questions in the survey, eight of which were mandatory. Overall, the completion rate for open-ended questions was high, with an average of 85\% for required questions and 82.9\% for non-mandatory ones. This indicates strong participant engagement, even when responses were optional. The quality of responses varied across questions, with some eliciting brief answers and others generating in-depth, detailed feedback.
Following designing and testing, our survey was distributed digitally using Google Forms. Participants were recruited using convenience sampling methods \cite{stratton2021population}, using social media platforms like WhatsApp and an email call within our institute. 
Before filling out the survey, we provided our participants with brief information about our study's aim, potential risks, benefits, and confidentiality policy, followed by an informed consent form. Our survey did not collect any identifiable information to maintain anonymity. We got 77 responses to our survey, out of which 75 participants filled out the complete form; the demographic details of our participants are provided in Table \ref{tab:Survey}. All the valid survey responses were exported into Google Sheets for analysis. We analyzed the closed-ended question (n=13) using descriptive analysis, such as calculating percentages, cross-tabulation, and visualizations. For open-ended questions (n=14), we performed thematic analysis \cite{clarke2017thematic} and generated codes such as \textit{"Self-reflective practices"}, \textit{"Challenges in Emotion Identification"}, and \textit{"Emotional Literacy"}. Examples of themes we identified were \textit{"Emotional Awareness and Regulation"} and \textit{"Language and Emotional Expression"}.


\begin{table*}[h]
\centering
\begin{tabular}{@{}ll@{}}
\toprule
\textbf{Category} & \textbf{Details and Count} \\ \midrule
\textbf{Age} & Range: \textbf{18 - 41}, Mean = \textbf{24.9}, SD = \textbf{3.54} \\
\textbf{Gender} & Males = \textbf{46}, Females = \textbf{28}, Prefer not to say = \textbf{1} \\
\textbf{Education} & Bachelors = \textbf{42}, Masters = \textbf{21}, Doctorate = \textbf{7}, Senior High School = \textbf{4}, Vocational Diploma = \textbf{1} \\
\textbf{Occupation} & Students = \textbf{22}, Professionals/Business = \textbf{24}, N/A = \textbf{29} \\ 
\textbf{Prior Experience} & No Experience = 38, With Experience = 37 (Journaling, Mindfulness, Self-reflection/Introspection,\\
& Apps and Wearables) \\
\bottomrule
\end{tabular}
\caption{Summary of Survey Participants' Demographics. Prior experience included details about participants' experience using tools and techniques for emotion tracking or management. Note: Participants mentioned more than one technique, and no experience means people do not actively track or manage emotions in their daily lives.}
\label{tab:Survey}
\end{table*}

\subsection{Interviews}

We conducted our formative semi-structured interviews with 32 participants. To guide our semi-structured interviews, we adopted the 5W1H framework \cite{yu2010study}. This framework was particularly well-suited for our study since it was an early-stage design study, which aimed at exploring how individuals perceive, approach, and reflect on the act of annotating or self-reporting emotions in their daily lives. As emotion tracking is a deeply personal and context-dependent practice, we needed a method that could surface not only what participants do, but also why and how they do it, within the broader landscape of their routines, motivations, and challenges. Prior to conducting our participants' interviews, we did pilots with 5 participants to understand the flow of our interview design and the relevance of questions. Our interview design was further guided by prior qualitative research on emotions \cite{nepal2024contextual, 10.1145/3699761, 10.1145/3699755}.
Below is an explanation of our interview design:
\textbf{1) "WHO}- are they?”: Focused on understanding participants’ emotional awareness, experiences, and familiarity with technology for emotion tracking. Questions explored how they perceive and manage emotions, their prior experience with emotion data collection and logging, and the perceived psychological impact of the process on their lifestyle. 
\textbf{2) "WHAT}- would they annotate?”: Examined the types of emotions or emotional events participants considered worth annotating. Questions included privacy concerns and whether they would share detailed information about their emotions. 
\textbf{3) "WHEN}- would they annotate?”: Addressed the timing and frequency of emotion annotation. Participants were asked about their preferences for real-time versus retrospective annotation, the contexts or scenarios where they felt annotation was most appropriate, what kind of prompts, and at what frequency they might prefer to be notified for annotating. 
\textbf{4) "WHERE}- would they annotate?”: Explored the environments where participants would feel comfortable annotating their emotions, as well as locations they might avoid. 
\textbf{5) "WHY}- would they annotate?”: Investigated participants’ motivations for annotating emotions, including perceived benefits and potential challenges or barriers. We asked them to discuss the benefits they see in annotating both positive and negative emotions.
\textbf{6) "HOW}- would they annotate?”: Delved into preferred tools and methods for annotation, the time participants were willing to dedicate, and their expectations for simplifying or improving the annotation process. A detailed description of our interview questions is provided in Appendix \ref{interviewapp}.

Our participant pool was well-educated, technology-friendly individuals aged 18 and above, with and without any experience of emotion tracking technology, recruited through convenience sampling \cite{stratton2021population}, using social media platforms like WhatsApp, as well as an email call. We received interest from 32 individuals for the interviews. Before conducting the interviews, we obtained digital consent from each participant through Google Forms sent via email. Along with their consent, we also collected information on their age, gender, education, current occupation, mental health conditions, prior experience with therapy/counseling, and wearables/emotion tracking/emotion data collection studies. Sixteen of our participants had prior experience with either participating in emotion data collection studies or using wearables for stress detection and emotion/mood tracking applications. The rest of our participants did not use technology-based mediums to understand their emotions and mostly relied on techniques such as self-introspection, meditation, exercises, communication with other people, or other mindfulness or coping techniques to deal with emotions. Details about our interview participants are summarized in Table \ref{tab:Interview}. 

\begin{table}[h]
\centering
\begin{tabular}{@{}ll@{}}
\toprule
\textbf{Category} & \textbf{Details and Count} \\ \midrule
\textbf{Age} & Range: \textbf{19 - 43}, Mean = \textbf{26.96}, SD = \textbf{7.15} \\
\textbf{Gender} & Males = \textbf{15}, Females = \textbf{17} \\
\textbf{Education} & High School/Diploma = \textbf{3}, Bachelors = \textbf{17}, Masters = \textbf{6}, Doctorate = \textbf{3}, Postdoctoral = \textbf{3}\\
\textbf{Occupation} & Students = \textbf{19}, Professionals/Business = \textbf{13} \\ 
\textbf{Attended Therapy} & Yes = \textbf{13}, No = \textbf{19} \\
\textbf{Diagnosis} & Yes = \textbf{2}, No = \textbf{30}\\
\textbf{Prior Experience} & No Experience = \textbf{16}, With Experience = \textbf{16} \\
\bottomrule
\end{tabular}
\caption{Summary of Interview Participants Demographics. Diagnosis includes details about participants' Mental Health Diagnosis. Prior experience included details about participants' experience of using tools and techniques for emotion/mood tracking and participating in emotion-data collection studies. No experience includes participants who don't actively use any technology to manage their emotions. Note: Participants mentioned more than one technique.}
\label{tab:Interview}
\end{table}

The interviews were conducted in English, either online or offline, based on the participant's preference. Each session was recorded using Zoom Pro, following verbal consent. The interviews began with a brief introduction to the study and familiarization with terms such as "emotions," “emotion annotations,” and “emotion intensity” to ensure participants understood the terminology and process of emotion data collection. The definition used to explain emotion annotation to our participant is \textit{"The process of identifying, labeling, and documenting emotional experiences, often to capture emotional data for research, self-reflection, or technological applications. It involves assigning labels (e.g., specific emotions like happiness, anger, or sadness) to emotional events using methods such as written records, mood-tracking apps, emojis, or voice recordings."} Each interview lasted approximately 30 minutes. The sessions were transcribed using Zoom’s built-in audio transcription feature. The transcribed documents were then exported to Google Docs and manually reviewed by the first and second authors for grammatical and transcription errors using the original voice recordings. Following transcription, we performed the inductive thematic analysis \cite{braun2006using}. We began with authors 1 and 2 reading and re-reading the interview to familiarize themselves with all the data individually. Following this, they individually generated the initial codes for all the data, which included codes like \textit{"Avoidance and denial as coping mechanism," "Understanding of basic emotions"}, and \textit{"Challenges with using static Likert scales"}. Later, authors 1 and 2 grouped similar codes together to form potential themes. Following this, all the authors reviewed the themes together by reviewing the data within each theme to ensure it accurately reflected the data. The high-level themes included \textit{"Emotional literacy and awareness," "Technology and concerns"}, \textit{"Emotional Regulation and management methods"}, and \textit{"Barriers to Identifying and Annotating Emotions"}. All authors reviewed and refined the themes iteratively to ensure they were coherent and distinct until saturation. The themes formulated in the process helped us to structure our findings (see section \ref{findings}). 


\begin{table}[h]
\centering
\begin{tabular}{@{}ll@{}}
\toprule
\textbf{Category} & \textbf{Details and Count} \\ \midrule
Professional Title & Psychologist = \textbf{1}, Clinical Psychologist = \textbf{2}, Psychiatrist = \textbf{6}, \\
& Peer Counselor = \textbf{3} \\
Year of Experience & Less than 1 year = \textbf{2}, 1-3 years = \textbf{4}, 4-7 years = \textbf{1}, \\
 & 8-10 years = \textbf{3}, More than 10 years = \textbf{2} \\
Experience with AI and Wearable Technology & No = \textbf{8}, Yes = \textbf{4} \\ \bottomrule
\end{tabular}
\caption{Summary of Focus Group Discussion Participants Demographics.}
\label{tab:professional_details}
\end{table}

\subsection{Focus Group Discussion}\label{fgd}
To conduct our focus group discussions (FGDs), we utilized purposive sampling \cite{tongco2007purposive} to recruit mental health professionals. We reached out to our collaborators, including doctors and NGOs, who helped disseminate our call for participation along with an interest form. From the 18 responses we received, 12 professionals were available for the scheduled time slots. We conducted three separate FGDs: FGD1 included 4 professionals (1 Psychiatrist and 3 Peer Counselors), FGD2 included 3 professionals (2 Clinical Psychologists and 1 Psychiatrist), and FGD3 included 5 professionals (1 Psychologist and 4 Psychiatrists). All participants in the focus groups were from the same country and shared a common cultural background as interview and survey participants, minimizing variability due to cross-cultural differences. Our FGD was designed in line with the prior qualitative studies done with domain experts \cite{10.1145/3699755, 10.1145/3631700.3664910, 10.1145/3411764.3445518}.
Prior to the FGDs, we obtained digital consent from each participant through Google Forms sent via email. In addition to consent, we collected information on their professional titles, years of experience in the mental health field, and familiarity with AI or wearable technology. Details about the participants are summarized in Table \ref{tab:professional_details}. All our FGDs were conducted online via Zoom Pro, with a single moderator leading each session following verbal consent to record the meeting. Each FGD began with a brief introduction of all participants within the discussion, followed by introduction slides presented by the moderator to outline the role of AI in healthcare, the definition of emotion AI, and a brief description of the current practices in AI for emotion data collection and labeling. This overview ensured that all participants had a shared understanding before the discussions began. Following this introduction, the discussion was organized into three main segments: 1) Current Practices for Assessing Emotional States, 2) Attitudes Towards Data and AI, and 3) Challenges and Opportunities in Emotion Data Collection and Recognition. In the first segment, professionals discussed their current methods for assessing the emotional states of patients and clients. The second segment focused on their initial impressions of using AI to understand and monitor emotions, including potential benefits and drawbacks in clinical settings. In the final segment, participants reviewed the current practices of AI data collection, as described in the introduction, and provided their perspectives, recommendations, and insights based on their own practices for the future. A more detailed description of our FGD is provided in the appendix \ref{FGDAPP}.
Each FGD lasted approximately 1 hour and 15 minutes and was conducted in English. The sessions were transcribed using Zoom's built-in audio transcription feature. The transcribed documents were then exported to Google Docs and manually reviewed by authors for grammatical and transcription errors using the original voice recordings. The first two authors then completed the familiarization, where they thoroughly read all the transcripts. Next, initial codes are generated by reading all the data systematically, highlighting data segments, and assigning brief labels that capture their essence, following inductive thematic analysis \cite{braun2006using}. The initial codes included \textit{"Positive attitude towards AI"} and \textit{"Emotional labeling is a mix of subjective and objective labels"}. The authors 1 and 2 searched for themes by grouping similar codes, forming broader patterns. Following this, all the authors jointly reviewed and refined the themes to generate coherent and distinct themes for structuring the findings (see section \ref{findings}). A few examples of our identified themes are \textit{"Parallel source of information"} and \textit{"Emotional ground truth"}.


\subsection{Development of Guidelines}

To develop our guidelines, we analyzed the data from each source (surveys, interviews, and focus groups) independently to identify recurring themes and patterns. Next, we grouped similar themes and organized them iteratively \cite{braun2006using} under the three guideline stages \textit{"Pre-data collection," "During data collection," and "Post-data collection"}, ensuring logical flow and coherence. We then cross-referenced our identified themes with methodologies and recommendations from prior studies to validate and expand our understanding \cite{8730884, 9779458, 10.1145/3025453.3025750, larradet2020toward, 10.1038/s41597-021-00945-4, kang2023k}. Finally, we synthesized the insights from participant data and literature to create an end-to-end framework for everyday emotion data collection that is practical, evidence-based, and user-centered. This process resulted in 15 guidelines (named G\#) divided into three data-collection stages, as presented in Table \ref{Guideline1}, \ref{Guideline2}, and \ref{Guideline3}. Further, we evaluated our guidelines for their validity in emotion AI research (see section \ref{eval}). 


\section{Designing \textit{AnnoSense} - An Everyday Emotion Data Collection Framework for AI}\label{findings}

This section introduces \textit{AnnoSense}, a framework comprising 15 guidelines designed to support robust emotion data collection in everyday contexts, enabling the development of AI models applicable to real-life scenarios. \textit{AnnoSense} is structured into three phases: pre-data collection, during-data collection, and post-data collection. Within each subsection, we will present findings from our data surveys, interviews, and FGDs to demonstrate the data-driven origins of each guideline. To ensure clarity and ease of navigation, we begin by presenting our data observations, structured into thematic subsections. These are followed by a dedicated subsection—Derived Guidelines—which outlines design guidelines that are directly informed by and grounded in the preceding observations.

\begin{table}[h!]
\centering
\begin{tabular}{llcc}
\hline
\textbf{Survey Question} & \textbf{Response} & \textbf{Count} & \textbf{Percentage} \\
\hline
\multirow{5}{*}{\shortstack[l]{How often do you take time to\\ reflect on your emotions?}} 
    & Never     & 2  & 2.7\%  \\
    & Rarely    & 11 & 14.7\% \\
    & \textbf{Sometimes} & \textbf{27} & \textbf{36.0\%} \\
    & Often     & 25 & 33.3\% \\
    & Always    & 10 & 13.3\% \\
\hline
\multirow{5}{*}{\shortstack[l]{When experiencing a strong emotion,\\ how easily can you identify the emotion?}} 
    & Very difficult & 2  & 2.7\%  \\
    & Difficult      & 12 & 16.0\% \\
    & Neutral        & 9  & 12.0\% \\
    & \textbf{Easy}  & \textbf{40} & \textbf{53.3\%} \\
    & Very easy      & 12 & 16.0\% \\
\hline
\multirow{5}{*}{\shortstack[l]{How often do you feel mixed emotions\\ (experiencing multiple emotions at once)?}} 
    & Never     & 1  & 1.3\%  \\
    & Rarely    & 18 & 24.0\% \\
    & \textbf{Sometimes} & \textbf{30} & \textbf{40.0\%} \\
    & Often     & 24 & 32.0\% \\
    & Always    & 2  & 2.7\%  \\
\hline
\end{tabular}
\caption{Participant responses on emotional awareness and reflection (N = 75)}
\label{tab:emotional_awareness_percent}
\end{table}

\subsection{\textit{"Two-way Communication"}: Pre-Data Collection Phase (G1-G6)}\label{Pre}

\subsubsection{\textbf{Data Observations:}} To design our pre-study guidelines, we analyzed data from surveys, participant interviews, and focus group discussions to understand participants' specific needs prior to data collection. 

\begin{figure*}[t]
    \centering
    \begin{subfigure}[b]{0.48\textwidth}
        \centering
        \includegraphics[width=\textwidth]{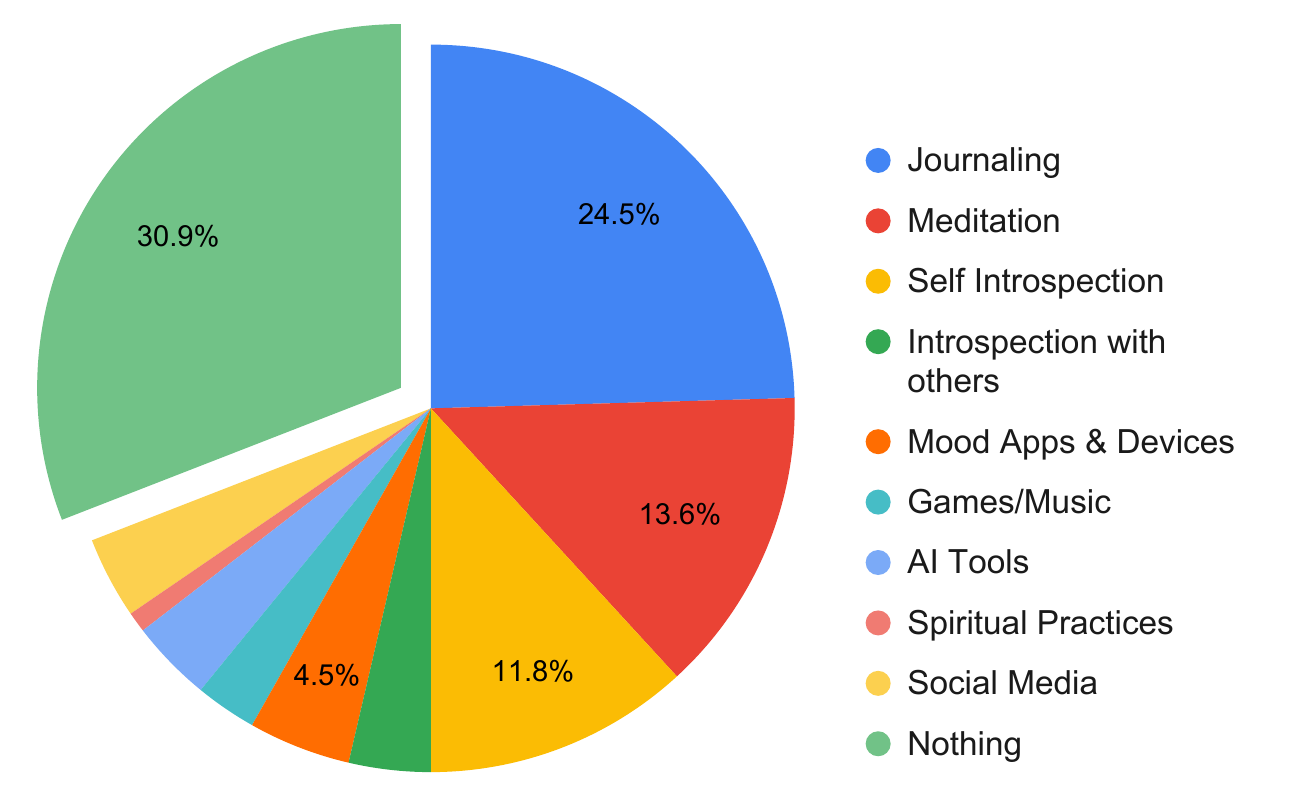}
        \caption{Survey results on emotion tracking practices.}
        \label{fig:emotion-pra}
    \end{subfigure}
    \hfill
    \begin{subfigure}[b]{0.48\textwidth}
        \centering
        \includegraphics[width=\textwidth]{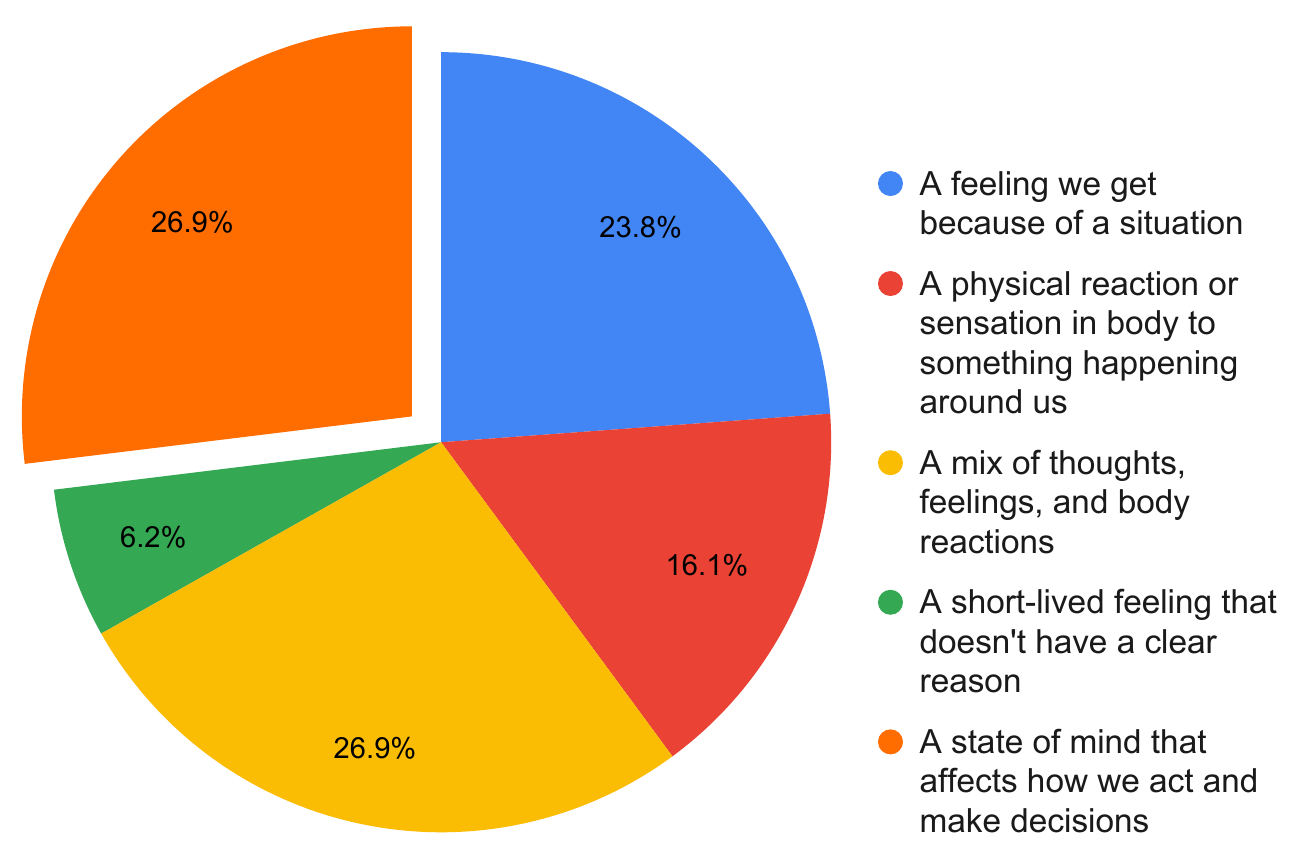}
        \caption{Participants' understanding of the definition of emotion.}
        \label{fig:define-emotion}
    \end{subfigure}
    \caption{Survey results for emotion awareness and management practices among our participants.}
\end{figure*}

\textbf{1) Need for Prior Preparation and Training:} Our survey findings indicate that participants generally believed that they possess moderate to high emotional awareness, with 69.3\% reporting that they can easily identify their emotions during intense emotional experiences. However, their emotional reflection habits vary considerably, 46.6\% reported to engage consistently in self-reflection, while a notable portion rarely or never does (more details in Table \ref{tab:emotional_awareness_percent}). Although many participants acknowledge experiencing mixed emotions, suggesting an awareness of emotional complexity, fewer than half use structured methods such as journaling, meditation, or introspection to process these feelings (see Figure \ref{fig:emotion-pra}). A significant number (34 participants) reported using nothing at all or using suppression or avoidance techniques such as social media and video games, implying that emotional insight for many relies on instinct rather than intentional strategies. This trend was further observed in our interview participants. Participants reported using distraction techniques like watching movies, playing games, or avoiding emotions as a common method for dealing with emotions (mostly negative).  As expressed by \textbf{(P8, Interview)} - \textit{"I usually sleep. I usually watch television. I usually watch a web series. Nothing else means I can do anything, or I just play some video game. That is the only way of dealing with these emotions, like, sometimes when I'm too stressed"}. Moreover, participants also mentioned not talking about or reflecting on deeper negative emotions due to the stigma of sharing or acknowledging emotions. Participants have used statements like - \textit{"Emotions make me feel weak"}, \textit{"It's better to keep emotions inside"}, or \textit{"Why should we track emotions? It is for people with mental disorders"} suggesting the deep-rooted stigma towards expressing, processing, or tracking emotions.

Furthermore, in our survey, when we asked participants to \textit{recall a recent situation in which they experienced a strong emotion and describe both the context and the emotion identified}, to explore participants’ emotional awareness, articulation, and the types of emotional experiences they tend to recall. A majority of participants (60\% ) were able to identify specific emotions tied to their experiences. Among these, anger, sadness, and anxiety were the most commonly reported emotions, suggesting negative emotions are commonly recalled by people. Mixed emotions were a notable part of the responses (9\%), reflecting the complex nature of human emotions. Lastly, 21\% of participants displayed uncertainty in expressing or identifying their emotions, with some responses showing emotional ambiguity or no clear emotion at all. This reflected the differences in participants' recall behaviors, where a majority of participants recalled negative events or were uncertain about expressing their emotions, possibly due to subconscious stigma or lack of vocabulary or awareness. 

Further, in our survey data, we found that a large number of participants are only aware of basic emotions such as happiness (46), sadness (28), joy (26), and anger (27), as shown in Table \ref{posneg}. Interestingly, these primary emotions, such as anger (20), happiness (18), and sadness (13), were also frequently reported as easily identifiable (see Table \ref{tab:emotioneasyhard}). We also observed that several emotions appeared in both "easy" and "hard" to identify categories such as, anger (20 vs. 10), sadness (13 vs. 8), anxiety (4 vs. 8), and happiness (18 vs. 4). This contradiction suggests the presence of distinct subgroups with varying levels of emotional literacy within our sample. Additionally, our survey data also revealed varying understanding among our participants about what they consider \textit{emotions}, as shown in Figure \ref{fig:define-emotion}.

\begin{table}[h!]
\centering
\begin{tabular}{llll}
\toprule
\textbf{Positive Emotion} & \textbf{Frequency} & \textbf{Negative Emotion} & \textbf{Frequency} \\
\midrule
Happy         & 46 & Sadness       & 28 \\
Gratitude     & 38 & Anger         & 27 \\
Joy           & 26 & Anxiety       & 16 \\
Hope          & 21 & Frustration   & 10 \\
Love          & 18 & \textbf{Motivation\textsuperscript{**}} & 9 \\
Peace         & 16 & Fear          & 9 \\
Excitement    & 14 & Loneliness    & 8 \\
Satisfaction  & 11 & Guilt         & 6 \\
Confidence    & 10 & Jealousy      & 6 \\
Motivation    & 9  & Stress        & 4 \\
Calm          & 9  & Irritation    & 4 \\
\bottomrule
\end{tabular}
\caption{Frequency of Top 10 Positive and Negative Emotions in Daily Life as Reported in our Survey. For positive emotions, the mean frequency of responses was 3.88 with a standard deviation of 1.8, while for negative emotions, the mean frequency was 2.48 with a standard deviation of 1.9. \textsuperscript{**}Note: \textbf{Motivation} is contextually a positive emotion but was mentioned in the negative list—possibly reflecting low or lack of motivation.}
\smallskip
\label{posneg}
\end{table}

\begin{table}[htbp]
    \centering
    \begin{tabular}{lc|lc}
        \hline
        \multicolumn{2}{c}{\textbf{Emotions Easy to Identify}} & \multicolumn{2}{c}{\textbf{Emotions Hard to Identify}} \\
        \hline
        \textbf{Emotion} & \textbf{Count} & \textbf{Emotion} & \textbf{Count} \\
        \hline
        Anger & 20 & Anger & 10 \\
        Happiness & 18 & Sadness & 8 \\
        Sadness & 13 & Anxiety & 8 \\
        Frustration & 6 & Satisfaction & 4 \\
        Joy & 6 & Fear & 4 \\
        Anxiety & 4 & Happiness & 4 \\
        Love & 4 & Depression & 4 \\
        Loneliness & 3 & Positive & 3 \\
        Disappointment & 3 & Jealousy & 3 \\
        Hope & 3 & Guilt & 3 \\
        \hline
    \end{tabular}
    \caption{Comparison of top 10 emotions based on ease of identification as per our survey response.}
    \label{tab:emotioneasyhard}
\end{table}

\begin{table*}[htp]
\centering
\begin{tabular}{cp{0.92\textwidth}}
\hline
\textbf{Guideline} & \textbf{Description} \\
\hline
\textbf{G1} & \textbf{Selecting Participants}: \\
 & \textbf{1.} Clearly document the inclusion and exclusion criteria based on the study’s objective and data requirements. \\
 & \textbf{2.} Recruit individuals from diverse demographic groups (age, gender, culture, education, and occupation) in line with the study’s inclusion criteria and data-requirements. \\
 & \textbf{3.} Screen participants for alexithymia (difficulty identifying and expressing emotions) using standardized screening tools such as the Toronto Alexithymia Scale \cite{haviland1996structure} or Perth Alexithymia Questionnaire \cite{preece2018psychometric}, neurological disorders (e.g., cognitive impairments), and health conditions (e.g., cardiovascular issues or chronic illnesses) that might impact physiological signals, ability to identify and express emotions, and in line with the study's exclusion criteria. \\
\hline
\textbf{G2} & \textbf{Obtaining Informed Consent}: \\
 & \textbf{1.} Clearly explain the purpose, benefits, potential risks, compensation, voluntary participation, time commitment, and key procedures of the study in simple, accessible language. Provide enough information to the participants
without revealing details that could compromise the integrity of the study design. \\
& \textbf{2.} Clearly outline ethical approval and privacy measures, such as how participant data will be anonymized (e.g., removal of personal identifiers), compliance with relevant data protection laws (e.g., GDPR, HIPAA), secure data storage practices, and data sharing in the consent document. \\
\hline
\textbf{G3} & \textbf{Conduct Initial Calibration}: \\
 & \textbf{1.} Conduct a baseline session or calibration trials (as per study requirements and resources) to familiarize participants with the devices being used in the study. \\
 & \textbf{2.} Provide clear instructions on how to correctly wear the devices and ensure they are functioning accurately during data collection. \\
\hline
\textbf{G4} & \textbf{Provide Participant Training}: \\
 & \textbf{1.} Organize practice sessions where participants label their emotions (e.g., joy, anger, sadness), identify subtle distinctions (e.g., frustration vs. irritation), and document contextual factors such as environment, social interactions, and cultural norms in real-time or respond to controlled stimuli, enabling researchers to clarify doubts and improve annotation accuracy. \\
 & \textbf{2.} Educate participants about the broader impacts of emotion annotations and the data privacy measures in place to build initial trust and engagement.\\
 & \textbf{3.} Offer expert-verified resources, including video clips, audio recordings, or books, to improve participants’ emotional literacy and understanding of what is meant by emotion annotations. \\
\hline
\textbf{G5} & \textbf{Perform Detailed Psycho-Social Profiling of Participants}: \\
 & \textbf{1.} Collaborate with domain experts to collect detailed histories on emotional characteristics such as emotional range (the spectrum of emotions a person can experience, express, and recognize), emotional congruency (the consistency between inner feelings and outward expressions), emotional intensity (degree of an emotional experience), and emotional reactivity (the intensity and speed of an individual’s emotional response to a stimulus), and emotional vocabulary. \\
 & \textbf{2.} Collect contextual details such as past traumatic experiences, daily routines, work-life balance, family dynamics, emotional awareness, regulation habits, and potential stigma using standardized questionnaires or with the assistance of domain experts. \\
\hline
\textbf{G6} & \textbf{Collect Comprehensive Demographic and Medical Data}: \\
 & \textbf{1.} Gather detailed demographic information (e.g., age, gender, education, socio-economic status, personality traits, and medical information) based on the specific needs of your research questions. \\
 & \textbf{2.} Ensure that demographic data is relevant to the study objectives, is ethically approved, and captures any additional factors that may influence emotional responses. \\
\hline
\end{tabular}
\caption{Guidelines for Pre-Data Collection Phase}
\label{Guideline1}
\end{table*}

\textbf{2) Understanding the Participant Profile:} Extending our investigation into emotional literacy, analysis of our survey question \textit{"Can you differentiate between similar emotions (e.g., sadness vs. disappointment)?"} revealed significant variations in participants' emotional granularity capabilities. Results showed that 44.0\% of participants explicitly reported difficulty differentiating between similar emotions, while only 20.9\% indicated confidence in their ability to distinguish nuanced emotional states. The remaining 35.2\% provided responses that were difficult to categorize definitively. Further, in our data, we observed several recurring themes: 1) sadness was characterized as a broader mood and disappointment as a more targeted emotion, 2) disappointment was frequently framed as a response to unmet expectations, and 3) they were differentiated based on perceived control, emotional intensity, and temporal duration. These findings further reflected the varying emotional abilities among our participants, suggesting that a one-size-fits-all solution to emotion data collection might not be sufficient for collecting quality data. Further, our discussions with experts also reconfirmed the \textit{varying emotional literacy} as explained by an expert \textbf{(P1, Psychologist, FGD3)}, \textit{"People tend to feel only 4-5 basic emotions and lack a vocabulary to explain their emotions and must be taught...A therapist tries to teach people about emotional awareness to improve vocabulary as it helps them identify emotions more clearly, along with their professional methods."} To overcome these challenges, domain experts within our FGDs emphasized efficient history-taking to understand emotion data reliably. Further experts also emphasized the importance of assessing various emotional aspects such as \textit{emotional vocabulary}, \textit{emotional range} (the spectrum of emotions a person can experience, express, and recognize), \textit{emotional congruency} (the consistency between inner feelings and outward expressions), \textit{emotional intensity} (degree of an emotional experience), and \textit{emotional reactivity} (the intensity and speed of an individual's emotional response to a stimulus) to understand emotional data better. Finally, experts highlighted the importance of screening for conditions like alexithymia, which affects an individual’s ability to identify and describe emotions. Finally, our participants' data and experts' discussions also revealed that emotions are deeply personal, and participants will find it challenging to share emotional details without assurance of privacy.

\subsubsection{\textbf{Derived Guidelines:}} As reflected in our data, there were differences in participants' emotional awareness and attitude towards emotion management. This inspired our guidelines \textbf{G3, G4} on participant training and initial calibrations to ensure data collection methods are accessible and relevant to a diverse population. This is further crucial for collecting richer data. Discussions with domain experts further reinforced the need for participant training, and prior research has also shown that the varying ability in identifying and articulating emotions can adversely impact the quality of emotion data \cite{gao2023critiquing, 10.1145/3460418.3479338}. Further to mitigate these impacts, we have added guidelines \textbf{G5, G6}. These guidelines are necessary for collecting detailed additional information to effectively contextualize emotion data \cite{10.1145/3699761}. Further, keeping in mind the wide range of hypotheses that inspire emotion data collection, we have added \textbf{G1}, which also includes exclusion-inclusion and screening guidelines inspired by data-centric AI and prior emotion data collection \cite{10.1145/3571724, 10.1145/3699755, 9779458}. Diversity, necessary screening, alongside a comprehensive understanding of factors influencing emotional data, is crucial for ensuring the reliability of emotion assessments. Lastly, as expressed \textbf{(P3, Psychiatrist, FGD2)}, \textit{"...the issue will be regarding the privacy part, how the data is being stored by the AI ... And you know who has access to it and how it is being used by the 3rd party. So overall, there are a lot of privacy-related challenges because there will be a lot of sensitive information. How we are tackling this will be an important point. And it should be communicated early on."}, we have added \textbf{G2}. Elaborate informed consent was necessary alongside training and contextualization because stigma and privacy concerns, as visible in our data, can deeply influence the self-reporting behaviors. Consequently, together with our insights from data, alongside prior practices to collect quality data and emotion data collection methodologies, have informed our pre-dataset collection guidelines as provided in Table \ref{Guideline1}.

\begin{figure}[t]
    \centering
    \begin{subfigure}[b]{0.48\textwidth}
        \centering
        \includegraphics[width=\textwidth]{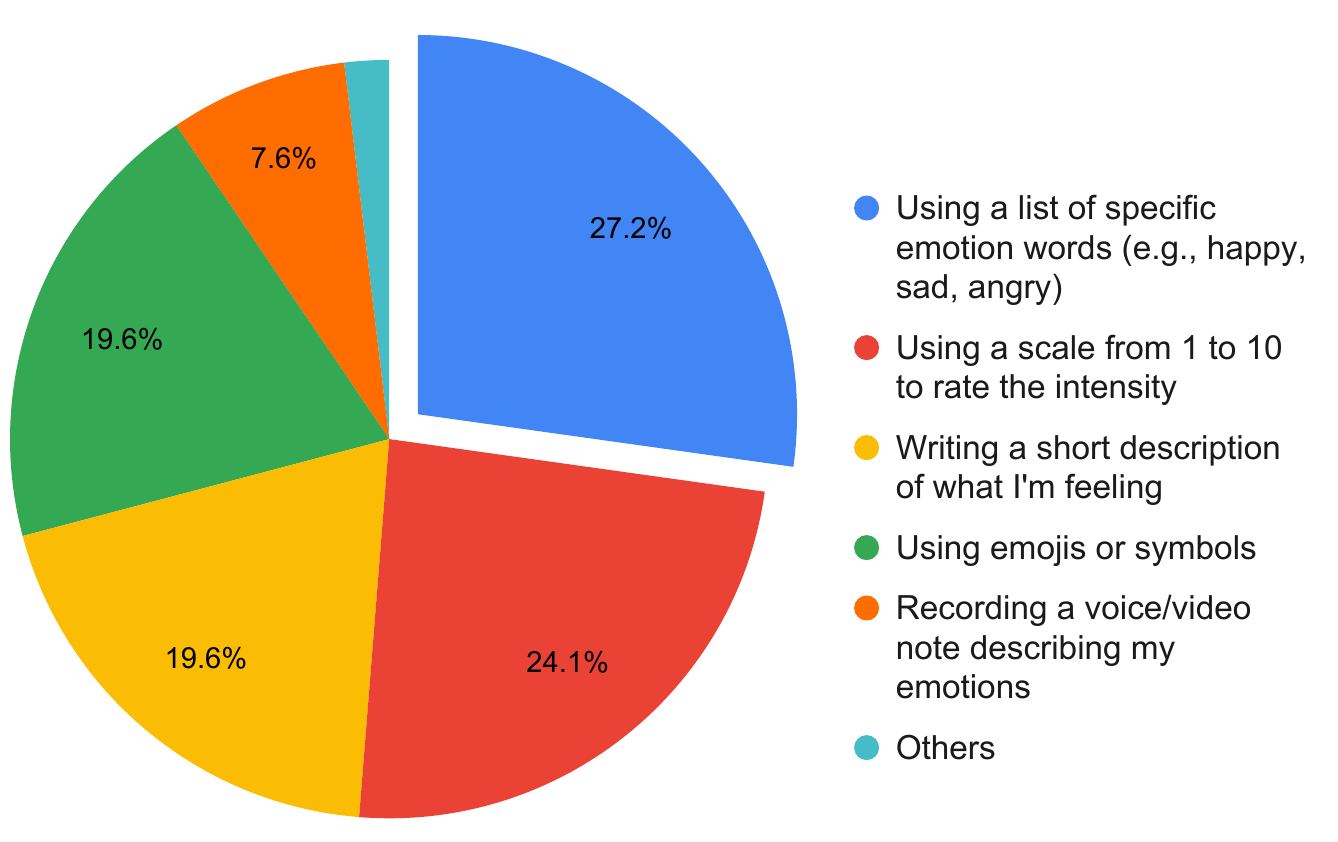}
        \caption{Preference of annotation method.}
        \label{fig:emotion-pref}
    \end{subfigure}
    \hfill
    \begin{subfigure}[b]{0.48\textwidth}
        \centering
        \includegraphics[width=\textwidth]{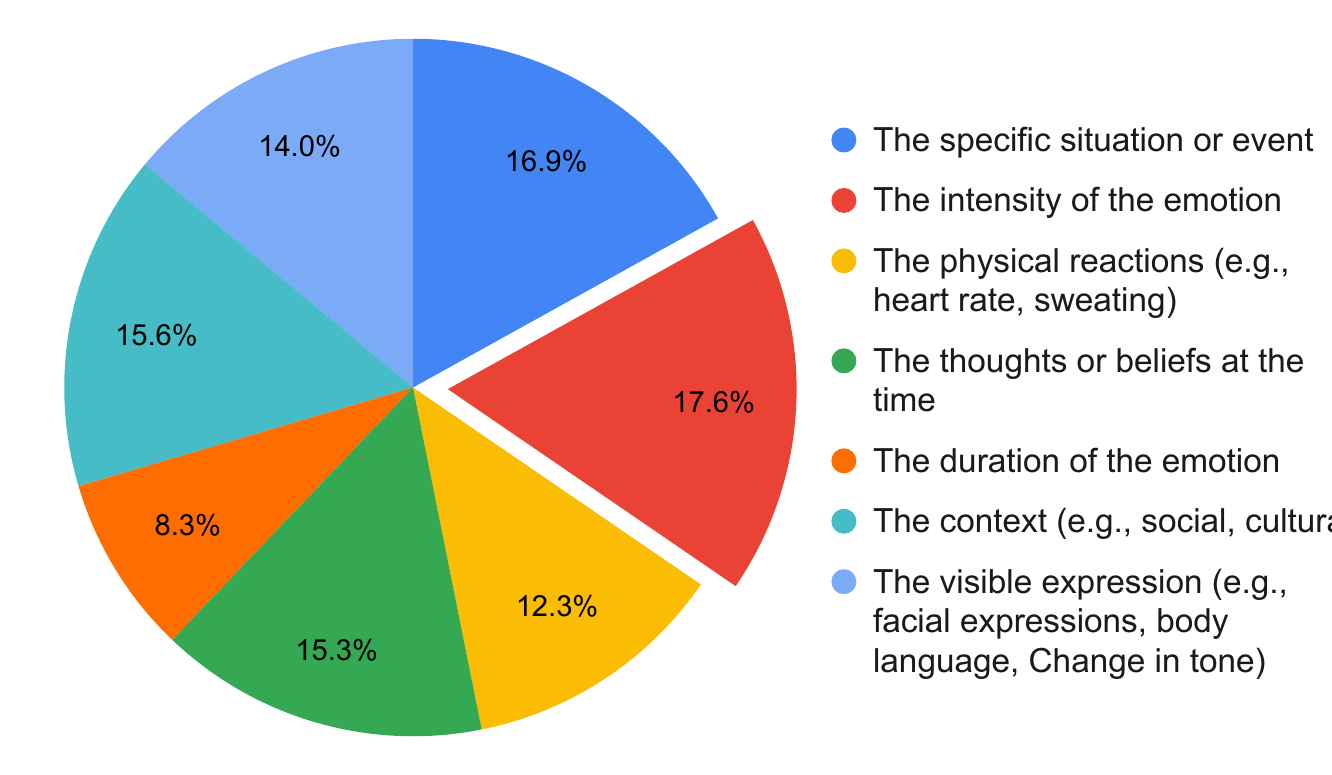}
        \caption{Factors that can impact annotations.}
        \label{fig:fact_ann}
    \end{subfigure}
    \caption{Survey results for attitude towards emotion annotation in daily life.}
\end{figure}

\subsection{Understanding the needs of "\textit{Data Source}": During Data Collection Phase (G7-G11)}\label{during}

\subsubsection{\textbf{Data Observations:}} Within everyday settings, participants are typically prompted to annotate their emotions based on random, fixed time, or event-based triggers in response to changes in physiological or activity data \cite{10.1145/3491102.3501944, 10.1038/s41597-021-00945-4, kang2023k}. These prompts often ask participants to fill out surveys or questionnaires based on pre-defined scales (as discussed in section \ref{datacollectionrw}). However, these predefined surveys offer limited flexibility for participants to share additional context or express emotions as per their intensities.

\textbf{1) Need for Adaptable Design}: In contrast to these rigid methods, our survey data highlighted the diverse preferences participants have when it comes to emotion annotation (see Figure \ref{fig:emotion-pref}). While 27.2\% of participants preferred using an emotion list, 19.6\% expressed a desire for an open-ended option to write about their emotions. Further, on examining participants' motivations behind their preferred annotation methods. \textit{Ease of use} emerged as the primary consideration (14.8\%), closely followed by \textit{expressiveness} (12.6\%) - the ability to fully convey emotional experiences. Participants also mentioned \textit{clarity} (8.9\% ) of methods and their ability to capture \textit{emotional complexity} (8.9\% ) as important factors. These findings suggest that participants are seeking annotation methods that balance accessibility with expressiveness, allowing them to capture nuanced emotional states. Moreover, the relatively even distribution across annotation preferences points to significant individual variation. Further, in our interview data, we found similar patterns that highlighted the need for annotation methods that consider the transient nature of emotions. As one participant \textbf{(P30 - Interview)} explained- \textit{"I would say the objective scales (likert, SAM or PANAS) will be easier use daily, but you should always give an option that if I am feeling extreme emotions — say if I'm extremely happy, extremely sad, or extremely angry —then there should be an option to write down or something."} Additionally, participants noted that during intense emotional moments, writing about the situation or their reactions would be easier rather than trying to identify and label specific emotions. 

This finding suggested that participants may struggle to articulate intense emotional experiences, highlighting the need for structured guidance to help them navigate and understand complex emotional states. Mental health experts reinforced this insight, recommending adaptable annotation methods modeled after diary writing approaches. They specifically suggested incorporating probing questions about triggers, situational contexts, and emotional reactions. Such structured frameworks can significantly reduce the cognitive burden of subjective emotion annotation, particularly for individuals experiencing complex or overwhelming emotional states. Further, our interview participants noted that emotions with visible cues are easier to identify. However, identifying and articulating complex emotions (such as co-occurring, mixed, layered, or ambiguous emotions) is challenging. These emotions—like anxiety combined with fear or frustration intertwined with anger—were described as harder to pinpoint. These insights highlight the importance of designing interfaces that can facilitate the expression of both simple and complex emotional experiences. Such an interface should provide participants with the required support, emotional vocabulary hints, reflective prompts, guided questions, and relatable analogies. 

A few participants also suggested using more abstract and expressive methods for annotating emotions. They felt that predefined scales or specific words often limited how they could express their feelings. Instead, they proposed alternatives like sharing the songs they were listening to, quotes that reflected their mood, or photos of their environment. Some also mentioned sketches or free-form journaling. These methods, as explained by participants, allowed for a more personal and authentic expression of emotions, reflecting the need for adaptable design.
In addition to this, our survey data also revealed varying factors that can influence the identification of emotions (see Figure \ref{fig:fact_ann}). The intensity of emotional experience as felt by a participant emerged as the most frequently mentioned factor (53 mentions), closely followed by the specific situational context triggering the emotion (51 mentions). Contextual elements, including social or cultural factors, thoughts present during emotional episodes, physical sensations, and visible expressions (facial expressions, body language, vocal changes) are also mentioned as crucial. This even distribution of factors further reinforced that participants recognize emotion as a multifaceted phenomenon requiring multidimensional annotation approaches.

\begin{table*}[htbp]
    \centering
    \begin{tabular}{>{\raggedright\arraybackslash}m{0.52\textwidth} 
                    >{\centering\arraybackslash}m{0.18\textwidth} 
                    >{\centering\arraybackslash}m{0.18\textwidth}}
        \toprule
        \textbf{Question} & \textbf{Response} & \textbf{Count} \\
        \midrule
        \multirow{2}{=}{Would you like to annotate your emotions daily?} 
            & Yes & 28 \\
            & \textbf{No} & \textbf{47} \\
        \midrule
        \multirow{5}{=}{How easy do you find it to annotate your emotions daily?} 
            & Very difficult & 5 \\
            & Difficult & 22 \\
            & \textbf{Neutral} & \textbf{29} \\
            & Easy & 16 \\
            & Very easy & 3 \\
        \midrule
        \multirow{6}{=}{How frequently can you annotate your emotions?} 
            & Multiple times a day & 17 \\
            & Once a day & 10 \\
            & \textbf{Few times a week} & \textbf{20} \\
            & Once a week & 11 \\
            & Less than once a week & 9 \\
            & Never & 8 \\
        \midrule
        \multirow{3}{=}{If you are going through a negative emotion, will you annotate?} 
            & \textbf{Yes} & \textbf{25} \\
            & No & 14 \\
            & Not Answered & 36 \\
        \midrule
        \multirow{3}{=}{If you are going through a positive emotion, will you annotate?} 
            & \textbf{Yes} & \textbf{21} \\
            & No & 20 \\
            & Not Answered & 34 \\
        \midrule
        \multirow{2}{=}{Do cultural or societal factors influence how you perceive emotions?} 
            & \textbf{Yes} & \textbf{44} \\
            & No & 31 \\
        \bottomrule
    \end{tabular}
    \caption{Survey Results on Emotion Annotation Practices and Perceptions}
    \label{annotationsurvey}
\end{table*}

\textbf{2) Participant's Agency, Learning and Participant-Aware Sampling:} Our analysis of emotion annotation preferences and practices survey data (see Table \ref{annotationsurvey}) revealed significant resistance to daily emotion tracking, with 62.7\% of respondents indicating reluctance compared to 37.3\% expressing interest. This reluctance corresponds with perceived difficulty, as 36.0\% found emotion annotation difficult, while only 25.3\% considered it easy. Frequency data further reinforced these patterns, with only 35.9\% of participants willing to annotate emotions daily, while 37.3\% preferred weekly or less frequently.
Further analysis of open-ended data and interviews revealed that participants wanted an annotation method that would prompt them according to their emotional intensities and pace. They also mentioned that the method should provide them feedback, insights, and an opportunity to learn from their data. Further, they mentioned that methods that only collect data without any learning engagements might not motivate them to annotate frequently. Participants also highlighted the importance of well-timed prompting methods per their personalized schedules. As mentioned by a participant (P26 - Interview) who uses an emotion logging application, the app frequently sends notifications when he begins working, distracting him. As a result, although he is willing to use the tracking technology, but he often does not annotate. 

This suggests that the varying needs of participants and assumptions, such as prompting users when they are not in motion, might not hold for everyone. For instance, while some may find such prompts helpful during idle moments, others—like P26—may perceive them as intrusive, especially when they coincide with focused work sessions. Our interviews further highlighted that the timing and context of annotation must align with users’ emotional states and willingness to engage alongside other contextual data such as activity levels, behavioral cues, and physiological changes. Further, our data also revealed that many participants preferred non-digital alternatives, viewing digital tools as requiring extra effort and time. Participants highlighted the availability of real-life alternatives (such as writing with pen and paper, sketching, sitting in silence, playing sports, or talking to friends) as the reason behind their preferences. This highlights the importance of participant-aware interventions incorporating user-agency in design and adapting to individual routines and preferences, rather than relying on one-size-fits-all strategies. Further, our survey data also revealed a significant component of cultural and societal influence on emotions. On deeper analysis, we found that these influences are shaped by negative connotations about sharing or expressing emotions, or stigma, and can hinder unbiased and balanced annotations. This suggested a need for an emotional literacy component in data collection methods.

\begin{table*}[htp]
\centering
\begin{tabular}{cp{0.92\textwidth}}
\hline
\textbf{Guideline} & \textbf{Description}\\
\hline
\textbf{G7} & \textbf{Focus on Participant's Agency}: \\
 & \textbf{1.} Use lightweight, non-intrusive wearable devices to avoid disrupting daily activities. \\
 & \textbf{2.} Allow users to adjust the annotation frequency based on their preferences or schedules while ensuring a minimum frequency that balances data accuracy with preventing fatigue and disengagement. \\
 & \textbf{3.} Set realistic expectations for emotional changes as per the research objective, for example conditions like depression don't show significant daily fluctuations, so daily recordings may not be necessary. \\
\hline
\textbf{G8} & \textbf{Develop Participant-Aware Sampling}: \\
 & \textbf{1.} Trigger annotations by corroborating information on participants' characteristics (gathered in G5) such as, daily schedules, activity levels, physiological changes, and emotional profile while keeping G7 and research objectives in mind. \\
\hline
\textbf{G9} & \textbf{Design Adaptable Annotation Methods}: \\
 & \textbf{1.} Offer participants the choice between structured annotation methods (e.g., SAM, PANAS) that use scales and unstructured subjective annotation methods (e.g., text, audio, images) as per their emotional intensity. \\
 & \textbf{2.} For subjective annotations, adopt structured frameworks like the ABC model (Activating Event, Belief, and Consequence) to guide participants' responses. Alternatively, design LLM-based prompts \cite{10.1145/3699761, 10.1145/3613904.3642790} customized to align with participants' unique emotional traits, as identified in steps G5, to provide tailored guidance.\\
 & \textbf{3.} Provide participants with support in understanding complex emotions by offering tools such as emotion vocabulary lists, options to select multiple emotions simultaneously, visual aids like emotion wheels, reflective prompts, guided questions, and relatable scenarios or activity list to foster emotional clarity.\\
\hline
\textbf{G10} & \textbf{Incorporate Multi-Perspective Assessments}: \\
 & \textbf{1.} Collect assessments not only from the participants themselves but also from trusted individuals in their support system, such as family members, peers, or mental health professionals, based on the participant cohort and study requirements. For example, clinical populations may require multiple assessments, whereas healthy individuals might need fewer.\\
 & \textbf{2.} Integrate additional data streams, such as location, social media activity, phone usage, sleep patterns, and calendar events. \\
 & \textbf{3.} Allow participants to select who and what data streams can contribute to their data based on their comfort and preferences. \\
\hline
\textbf{G11} & \textbf{Focus on Participant Engagement, Learning and Support}: \\
 & \textbf{1.} Periodically reach out to participants to address any concerns, clarify expectations, motivate, and support.\\
 & \textbf{2.} Use UI designs and prompts to encourage reflection, show growth, and provide supportive feedback to maintain engagement. \\
 & \textbf{3.} Integrate interventions within the study that help participants enhance their emotional literacy over time. \\
 & \textbf{4.} Integrate prompts that encourage reflection on positive outcomes or gratitude to offset the potential negative impact of recording difficult emotions. Additionally, provide access to mental health resources or emotional support tools for participants who may experience distress from self-reporting.\\
\hline
\end{tabular}
\caption{Guidelines for During Data Collection Phase}
\label{Guideline2}
\end{table*}

\textbf{3) Multi-perspective Assessments:} Finally, our discussions with experts emphasized the importance of collecting emotional data from multiple sources, specifically for people with mental disorders or significant life events. They recommended combining self-reports with family members' input and regular evaluations by psychologists or psychiatrists. Emotional assessment is complex —even for professionals— so relying on a single source may lead to unreliable results. Experts also highlighted the value of integrating additional data streams. These included ecological activity data, social media behaviors, physiological signals, and other automated, objective measures. These sources can help complement and contextualize subjective self-reports. Experts also stressed the need for emotion assessment methods tailored to different groups. For the general population, tools should focus on promoting wellness and addressing everyday stressors.
In contrast, more in-depth emotional assessments and professional evaluations are critical for clinical populations or those facing significant life events to ensure accurate and meaningful insights. Further, our interview data also highlighted a set of participants who were skeptical about using technology for managing emotions and emphasized the need for a human touch. This finding reinforced the importance of incorporating multi-source assessment approaches, involvement of trusted people, and thoughtful data sharing mechanisms into emotion data collection methodologies. By integrating these elements, emotion tracking systems can complement rather than replace human interaction.


\subsubsection{\textbf{Derived Guidelines:}} Our analysis revealed a preference for a balanced approach to emotion annotation. Participants valued having the flexibility to choose between structured, scale-based methods and unstructured, journal-writing methods based on the intensity of their emotions. The suggested need for this flexibility in our data collection approaches guided our addition of guideline \textbf{G9}. Within our data, it was also evident that participants frequently linked their emotions to specific environmental or situational cues. For example, loneliness was associated with the absence of companionship, stress with workload, fear with significant life events, and joy with time spent with loved ones. This underscores the need for tools that allow participants to articulate emotions by connecting them to contextual factors \textbf{(G9.3)}. Further, the need for user-agency to personalize the prompts per their schedules and emotional spectrum is also highlighted. Thus, designing methods with interfaces that could balance user-agency and participant-burden with data needs would be essential, guiding our inclusion of \textbf{G7}. Our data also highlighted a need to move beyond the context-aware sampling \cite{10.1145/3123024.3124571, nepal2024contextual}, and adding a layer of participants' persona, cultural knowledge \cite{veselovsky2025localizedculturalknowledgeconserved} to sampling strategies \cite{10.1145/3699761}, as included in guideline \textbf{G8}. In addition to it, our discussion with experts and participants' interviews highlighted a need for adding multiple-perspective assessments and the option to multi-source data contribution \cite{10.1145/2968219.2968301} for improving the data quality. This supported our addition of guideline \textbf{G10}. Finally, our data observations suggested a need for better participant support and components for improving emotion literacy over time in our data collection strategies for better participant engagement, guiding the inclusion of \textbf{G11}. Our detailed during data collection guidelines are presented in Table \ref{Guideline2}.

\subsection{Learning from \textit{Dynamic} Data: Post-Data Collection Phase (G12-G15)}\label{post}

\subsubsection{\textbf{Data Observations:}} Following data collection, the post-processing stage involves several critical steps to ensure data usability and integrity. Our data highlighted several observations for the post-data collection stage.

\textbf{1) Consistent Best Practices:} The post-processing stage typically includes quality checks, consistent structuring, and preparation for data sharing to enable reproducibility and collaborative research \cite{10.1145/3571724}. While prior work highlights these best practices, data sharing remains inconsistent. For example, datasets such as WESAD \cite{10.1145/3242969.3242985}, EEVR \cite{singh2024eevr}, and ASCERTAIN \cite{10.1109/TAFFC.2016.2625250} provide not only the data but also baseline experiments to support emotion recognition research. In contrast, datasets like EMOGNITION \cite{saganowski2022emognition} and GReX \cite{bota2024real} focus solely on data release with quality checks, without offering baseline evaluations. While valuable, the absence of standardized benchmarks increases friction for downstream use and hinders fair comparisons across studies.

\textbf{2) Handling Dynamic Data:} Further analysis, as discussed in Section \ref{during}, revealed that participants have diverse needs and preferences regarding the sharing of their emotion data. Recognizing and addressing these needs is critical for fostering participant engagement and trust. Incorporating such preferences into data collection practices can enrich the resulting datasets, enabling the integration of information from multiple sources and annotation methods, including both structured and unstructured formats.
Effectively managing this dynamic emotion data requires the establishment of a standardized data pipeline. This pipeline should include procedures for identifying and handling missing values, inconsistencies, and artifacts that could compromise data quality. Given the heterogeneous nature of emotion data, validation must also extend to the annotation layer. This involves assessing the reliability of labels through cross-validation across various sources—such as physiological signals, self-reports, behavioral observations, and expert annotations where applicable. Such practices enhance the robustness and accuracy of the dataset, ultimately offering a more comprehensive and trustworthy representation of participants' emotional experiences. Normalization is another critical pre-processing step, particularly because individual differences, such as physiological signal ranges, emotional reactivity, or even environmental factors like temperature, can significantly influence emotional data. It is also important to document when and why normalization is applied to ensure transparency in the data processing steps. Normalization is crucial because it helps reduce bias and variation that could lead to inaccurate conclusions. 

\begin{table*}[htp]
\centering
\begin{tabular}{cp{0.92\textwidth}}
\hline
\textbf{Guideline} & \textbf{Description} \\
\hline
\textbf{G12} & \textbf{Secure Data Handling}: \\
 & \textbf{1.} Store data securely using encryption and anonymization techniques, adhering to ethical guidelines (e.g., GDPR, HIPAA). \\
 & \textbf{2.} Allow participants to request data reviews within a specified timeframe (e.g., 30 days), with researchers providing an overview instead of direct access to raw data to avoid misinterpretation and better confidentiality. Authorized researchers should handle deletion to maintain data security if deletion is requested. \\
 & \textbf{3.} Clearly communicate any limitations on data review or deletion, such as once data has been anonymized or aggregated for analysis. \\
\hline
\textbf{G13} & \textbf{Data Quality Validation and Pre-processing}: \\
 & \textbf{1.} Review datasets for missing values, artifacts, or inconsistencies. Depending on the study's goals and the extent of missing values, methods such as imputation, removal, or flagging of problematic data can be used to handle these issues. \\
 & \textbf{2.} Cross-validate multiple data-sources (if available, G9 and G10) to improve reliability. \\
 & \textbf{3.} Normalize data where individual differences (e.g., physiological ranges, personality traits) or environmental factors (e.g., time of day, activity) significantly affect results. Document when normalization is applied and why. \\
\hline
\textbf{G14} & \textbf{Holistically Analyzing and Grounding the Data}: \\
 & \textbf{1.} Combine qualitative insights (e.g., text-based descriptions) and quantitative data (e.g., scale-based measures) to create multi-dimensional emotion labels that accurately capture the emotional experience within its context. \\
 & \textbf{2.} Ground data on the reliability and relevance of the source (if G10 is applicable), such as expert assessments for emotional dysregulation, peer feedback for social interactions, and self-reports for subjective experiences. \\
 & \textbf{3.} Combine psychosocial details (e.g., emotional traits, past experiences, daily routines) with emotion data to create a context-rich foundation for analysis and labeling, and document how these psychosocial factors impact emotion labeling to enhance transparency. \\
 & \textbf{4.} Collaborate with domain experts to review and ensure the accuracy and consistency of grounded emotion labels. \\
\hline
\textbf{G15} & \textbf{Share Findings, Best Practices, Data Limitations and  Usability}: \\
 & \textbf{1.} Present key findings and any challenges faced during data collection, such as participant engagement issues, device inaccuracies, or contextual variability. Describe the study protocol in detail to ensure reproducibility. \\
 & \textbf{2.} Highlight data limitations such as device reliability, data quality concerns, participant biases, or issues with ecological validity. \\
 & \textbf{3.} Specify the intended AI applications for the data, like emotion detection, disorder diagnosis, or longitudinal tracking of emotional changes. Then, evaluate the data's suitability for each of these specific use cases.\\
\hline
\end{tabular}
\caption{Guidelines for Post-Data Collection Phase}
\label{Guideline3}
\end{table*}

\textbf{3) Grounding Emotion data:} Our participant data highlighted that adopting more dynamic annotation approaches during data collection will likely produce annotations that differ from the standardized, fixed-scale labels typically used. These annotations will be more nuanced, context-dependent, and reflect participants’ real-time emotional experiences. Domain experts supported this view. They recommended using both structured and unstructured data when labeling emotions. They emphasized that relying on just one type of data could miss essential nuances. When discussing emotional "ground truth," experts pointed out that exact accuracy is less critical than generating actionable insights. They stressed that aligning different data sources is a key indicator of accurate emotion labeling. Although these dynamic annotation approaches will capture richer and more authentic emotional experiences, they pose challenges for applying traditional AI models. This highlights the need to design newer systematic approaches to ground the emotion data.


\textbf{4) Secure Data Handling:} As discussed in Section \ref{Pre}, our participants have expressed a preference for keeping their emotions private or sharing them only with trusted individuals, such as family members or mental health professionals. Sharing emotional information with others or through technology was not the first choice for many participants unless it significantly impacted their lives. Moreover, our survey (see Table \ref{annotationsurvey}), 62.7\% of participants expressed reluctance to track their emotions using digital tools. Alongside time constraints and concerns about overthinking, emotional privacy emerged as a key reason for avoiding such tools. This highlights a strong stigma around tracking or sharing emotional data, as expressed by \textbf{(P21, Interview)}: \textit{"If I am in real distress and I really want to get myself treated or understand the depth of my emotions, then I might provide access to my journal.} This suggests the need to maintain data security post-data collection.

\subsubsection{\textbf{Derived Guidelines:}} To address the privacy concerns, it is crucial that data collectors ensure participants feel confident in the secure handling of their data. Participants must also have the option to review or request deletion of their data, as outlined in \textbf{G12}. This guideline is essential because ensuring participants’ trust is the foundation of ethical data collection, particularly when dealing with sensitive emotional information. By offering data review and deletion options, we respect participants' autonomy and privacy, addressing stigma and data misuse concerns. Subsequently, after data collection, validation of data quality \cite{10.1145/3571724} is an essential step, guiding the addition of \textbf{G13}. This includes support for heterogeneous data formats, standardized metadata schemas, and robust pre-processing pipelines that handle noise, missing values, and temporal inconsistencies.
Further, the need for contextually grounding the collected dynamic data informed our guideline \textbf{G14}, which emphasizes the importance of holistically analyzing and grounding data. This guideline ensures that emotion labels reflect the complexity of participants’ lived experiences rather than oversimplifying them for AI processing. Grounding the data involves assessing the reliability and relevance of its sources. 

Recognizing psychosocial individual differences through standardized tools \cite{nock2008emotion, lane1990levels} or through expert interpretations to contextualize emotional labels more accurately. Additionally, combining unstructured subjective responses (by quantifying them using either psycholinguistic analysis \cite{sert2023review} or expert feedback) with structured, scale-based data provides a more nuanced and actionable grounding approach. This ensures that emotion annotations are participant-centered, addressing individual emotional experiences while also enabling the extraction of meaningful and useful emotion labels.
Lastly, \textbf{G15} outlines the necessity of transparently presenting key findings alongside any challenges encountered during data collection, such as participant engagement issues, device inaccuracies, or contextual variability. By detailing these challenges, researchers offer transparency into the reliability and scope of the data, which is critical for ensuring the reproducibility and validity of the findings. Additionally, specifying the intended applications for the collected data—whether for emotion detection, disorder diagnosis, or tracking emotional changes over time—is crucial for guiding its use. Further, benchmarking and evaluating the data’s suitability for the downstream task is equally essential for the future applicability of the dataset. Finally, our guidelines for the post-data collection phase are presented in Table \ref{Guideline3}.

\begin{figure}[h!]
  \centering
  \begin{subfigure}[b]{0.48\textwidth}
    \includegraphics[width=1\textwidth]
    {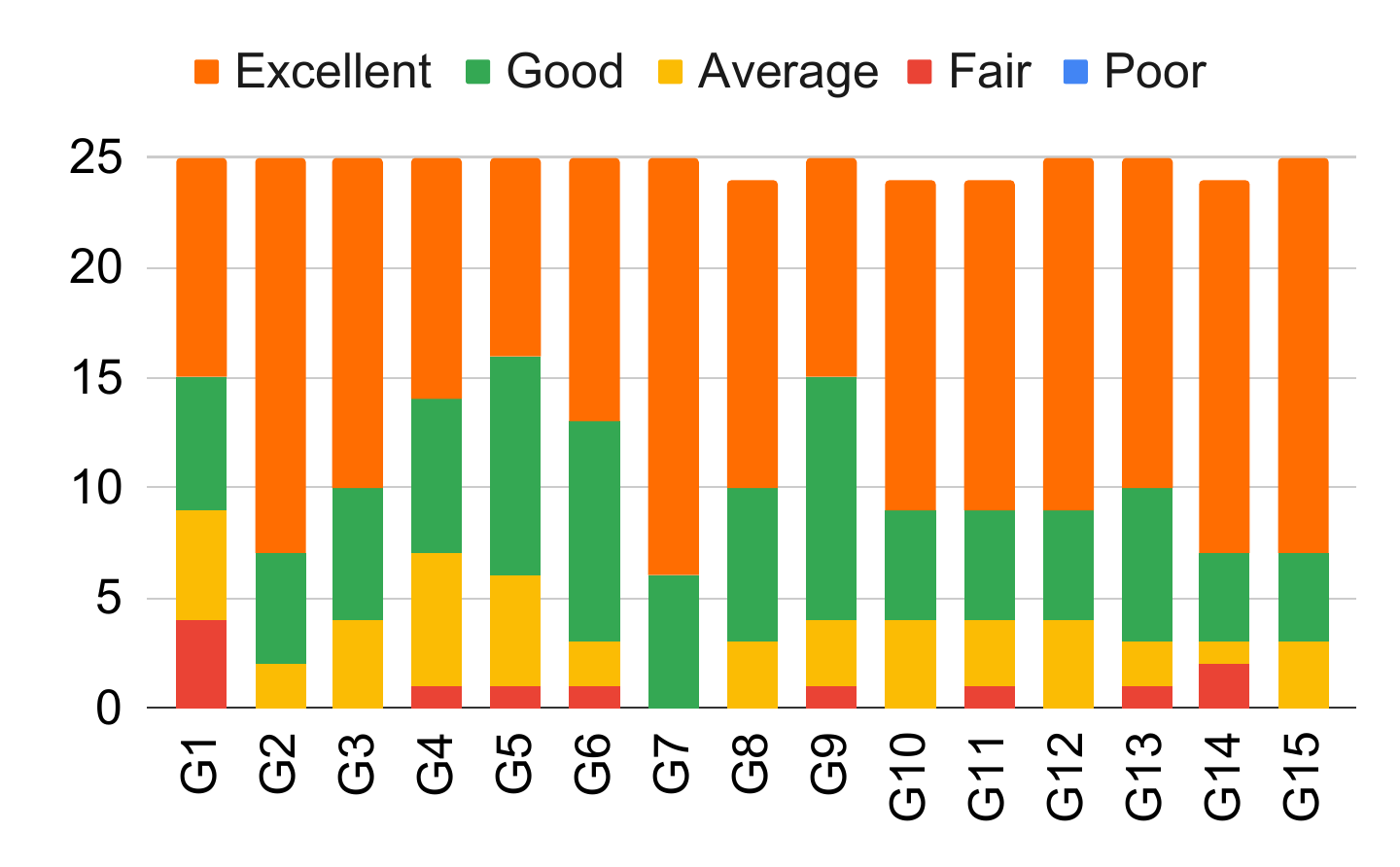} 
    \caption{Clarity}
    \label{fig:sub1}
  \end{subfigure}
  \hfill 
  \begin{subfigure}[b]{0.48\textwidth}
    \includegraphics[width=1\textwidth]{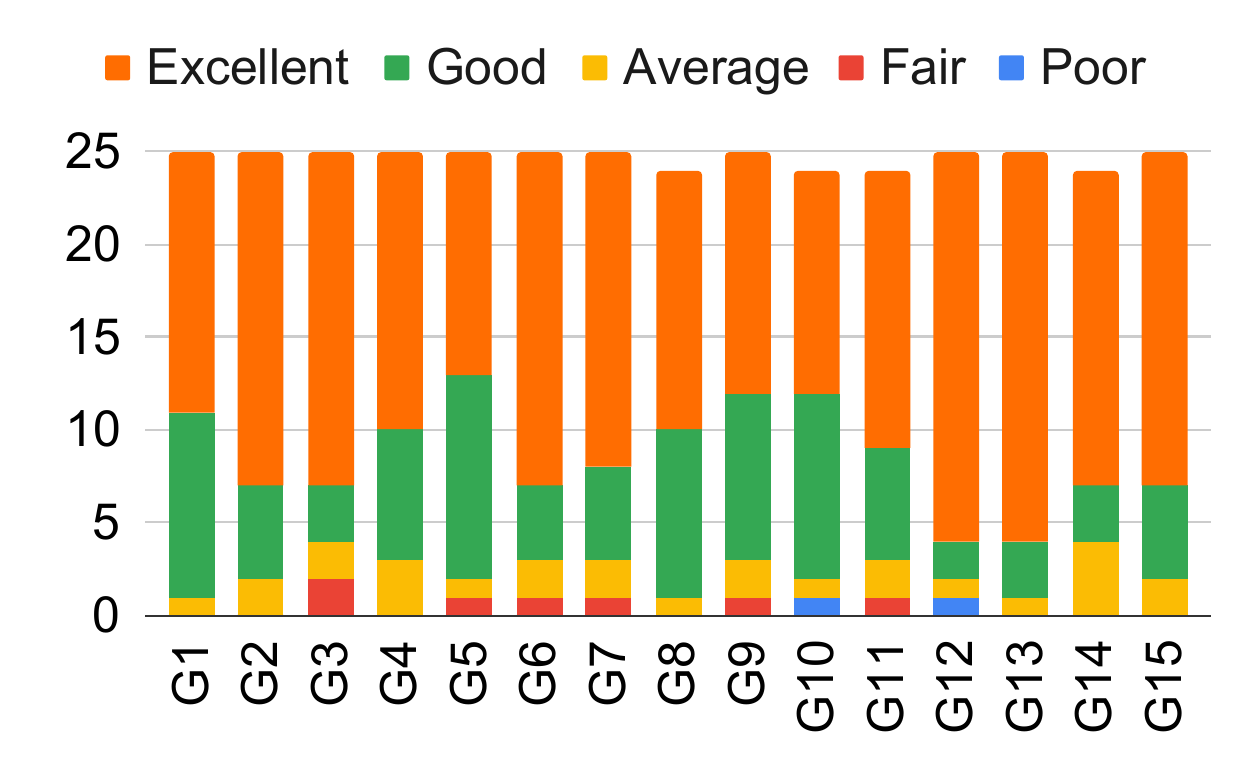} 
    \caption{Usefulness}
    \label{fig:sub2}
  \end{subfigure}
  \hfill 
  \begin{subfigure}[b]{0.48\textwidth}
    \includegraphics[width=1\textwidth]{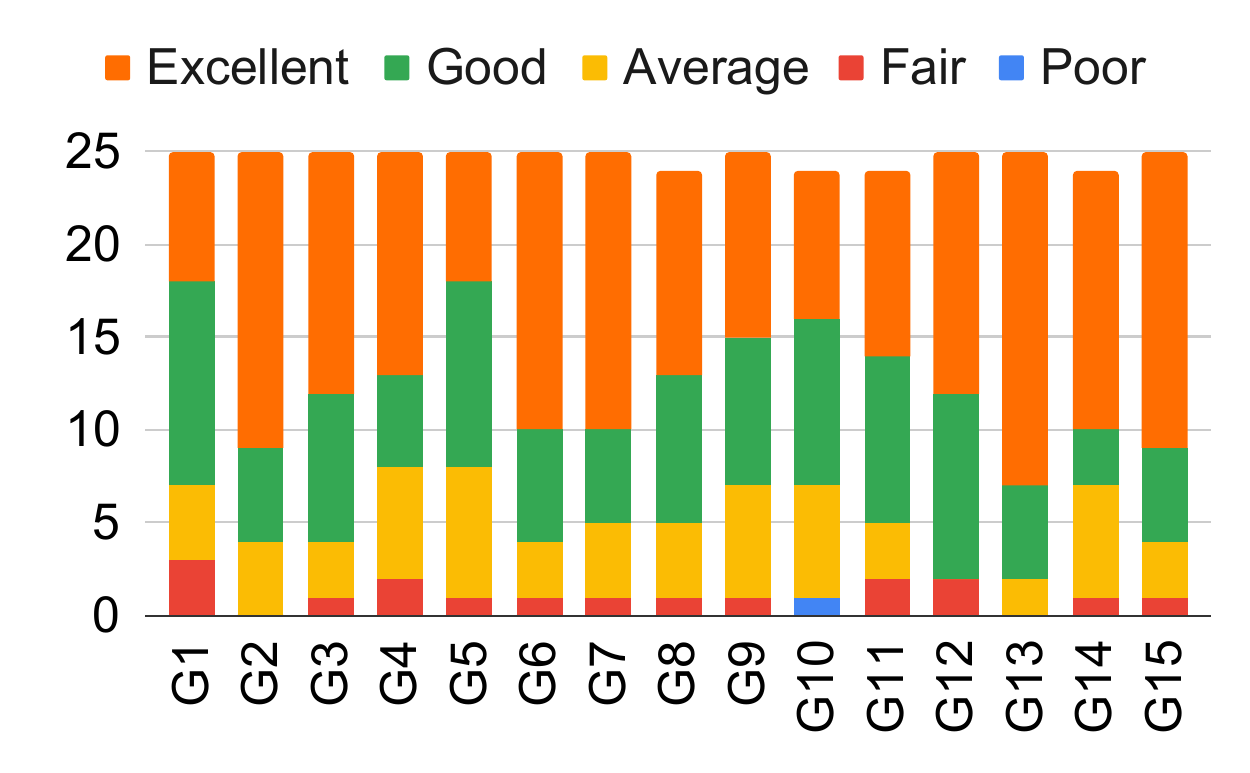} 
    \caption{Adaptability}
    \label{fig:sub3}
  \end{subfigure}
  
  \caption{The evaluator's scores for our guidelines.}
  \label{fig:main}
\end{figure}

\section{Evaluation of Guidelines}\label{eval}

The 15 guidelines were first internally evaluated by all the authors and fellow researchers/colleagues for clarity, usefulness, and adaptability. Following internal evaluations, we conducted an external evaluation involving 25 expert evaluators. These evaluators possessed expertise in emotion AI, physiological data collection, affective and ubiquitous computing research (detailed demographics presented in Table \ref{tab:guideline_details}). This evaluation aimed to gather expert feedback on the clarity, usefulness, and adaptability of our guidelines. We employed purposive sampling \cite{palinkas2015purposeful} to select these experts, who were then contacted via email and social media platforms such as WhatsApp, Twitter, and LinkedIn. Each expert received a survey form encompassing informed consent, demographic questions (area of expertise and years of experience), and the 15 guidelines themselves, organized into three sections: \textit{"Pre-data collection," "During data collection,"} and \textit{"Post-data collection"}. For each guideline, we asked the experts to provide their ratings for clarity (description and communication), usefulness (practicality and goal achievement), and adaptability (real-world applicability across diverse contexts) on a 5-point Likert scale (1=poor to 5=excellent). Additionally, we asked the experts to provide qualitative feedback in the form of comments/suggestions for further refining the guidelines. Our method draws inspiration from Amershi et al.'s modified heuristic evaluation \cite{amershi2019guidelines}, where we adapted the principles of discount usability testing to evaluate our guidelines. Furthermore, to refine the guidelines, we conducted a descriptive analysis \cite{loeb2017descriptive} of evaluator ratings as illustrated in figure \ref{fig:main}.

Descriptive analysis of the expert evaluations reveals a strongly positive overall reception of the guidelines across all assessed criteria - clarity, usefulness, and adaptability. The guidelines were consistently rated highly for their clarity and usefulness, with the vast majority of evaluators rating them as "Good" or "Excellent" in these domains. While Adaptability also received predominantly positive ratings, a slightly higher proportion of "Average" and "Fair" scores in this category suggests a potential for refinements to enhance their perceived applicability across diverse research contexts. Crucially, the consistent absence of "Poor" ratings across all guidelines and criteria indicated a robust framework without major perceived weaknesses. In addition to descriptive analysis, open-ended feedback was subjected to inductive thematic analysis \cite{braun2006using}, performed by the first author, to identify key suggestions for improvement. These suggestions, derived from expert feedback, primarily focused on enhancing clarity and comprehensiveness. Evaluators recommended adding more detailed explanations, illustrative examples, and definitions to make the guidelines more accessible. Furthermore, they emphasized the need to acknowledge the context-dependent nature of the guidelines, noting that their application may vary based on specific research objectives. In response to this feedback, we iteratively revised and rephrased the guidelines where needed to increase their adaptability to a wider everyday emotion research context. For example, Guidelines \textit{\#G1.1} and \textit{\#G1.2} were refined to explicitly state the importance of diverse recruitment while remaining aligned with specific study objectives. For \textit{\#G1.3}, to improve accessibility for interdisciplinary audiences, we incorporated references to screening tools like the Toronto Alexithymia Scale and added a definition of alexithymia. Similarly, Guidelines \textit{\#G3}, \textit{\#G4}, and \textit{\#G5} were revised to include examples and definitions, enhancing their overall clarity and broadening their applicability. Finally, all the authors then revisited and finalized the guidelines internally as presented in table \ref{Guideline1}, \ref{Guideline2}, and \ref{Guideline3}.

\begin{table}[h]
\centering
\begin{tabular}{@{}ll@{}}
\toprule
\textbf{Category} & \textbf{Details and Count} \\ \midrule
Gender & Male = \textbf{13}, Female = \textbf{12}\\
Year of Experience & 0-5 years = \textbf{12}, 5-10 years = \textbf{8}, 10+ years = \textbf{5}\\
Role & Researcher (Emotion AI/ Affective Computing/ HCI) = \textbf{23}, Data Scientist (Emotion AI) = \textbf{1}\\
& Researcher (Ubiquitous Computing/AI) = \textbf{1} \\ \bottomrule
\end{tabular}
\caption{Summary of Guidelines Evaluators.}
\label{tab:guideline_details}
\end{table}

\begin{figure*}[ht]
    \centering
    \includegraphics[width=1.1\textwidth]{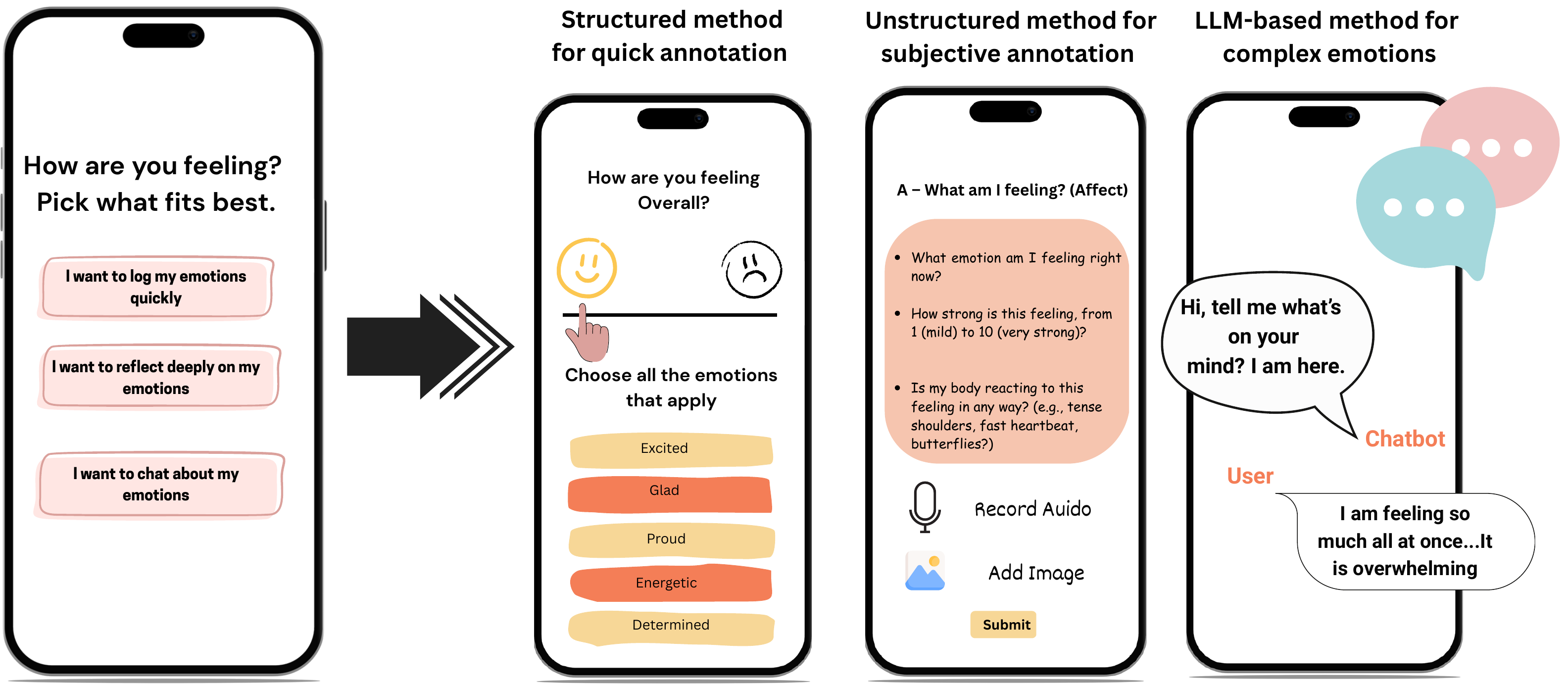} 
    \caption{Visualization of the Participant-Centric Adaptable Annotation Approach. This approach offers flexible annotation options tailored to participants’ emotional intensity and time availability. For quick annotations, a structured method using predefined emotion scales and lists is provided. In intense emotional experiences, participants can opt for a subjective, open-ended annotation guided by reflective questions. Additionally, large language model (LLM)-based support can facilitate meaningful annotation for users with lower emotional literacy.}
    \label{fig:annotation_visual1}
\end{figure*}

\section{Discussion}


This section discusses how future research can leverage \textit{AnnoSense} framework for designing participant-centric methodologies. First, we will discuss how to prototype tools based on our guidelines in section \ref{prototype}. Then, we will discuss the implications of pre, during, and post-data collection guidelines for future work.

\begin{figure*}[ht]
    \centering
    \includegraphics[width=1.1\textwidth]{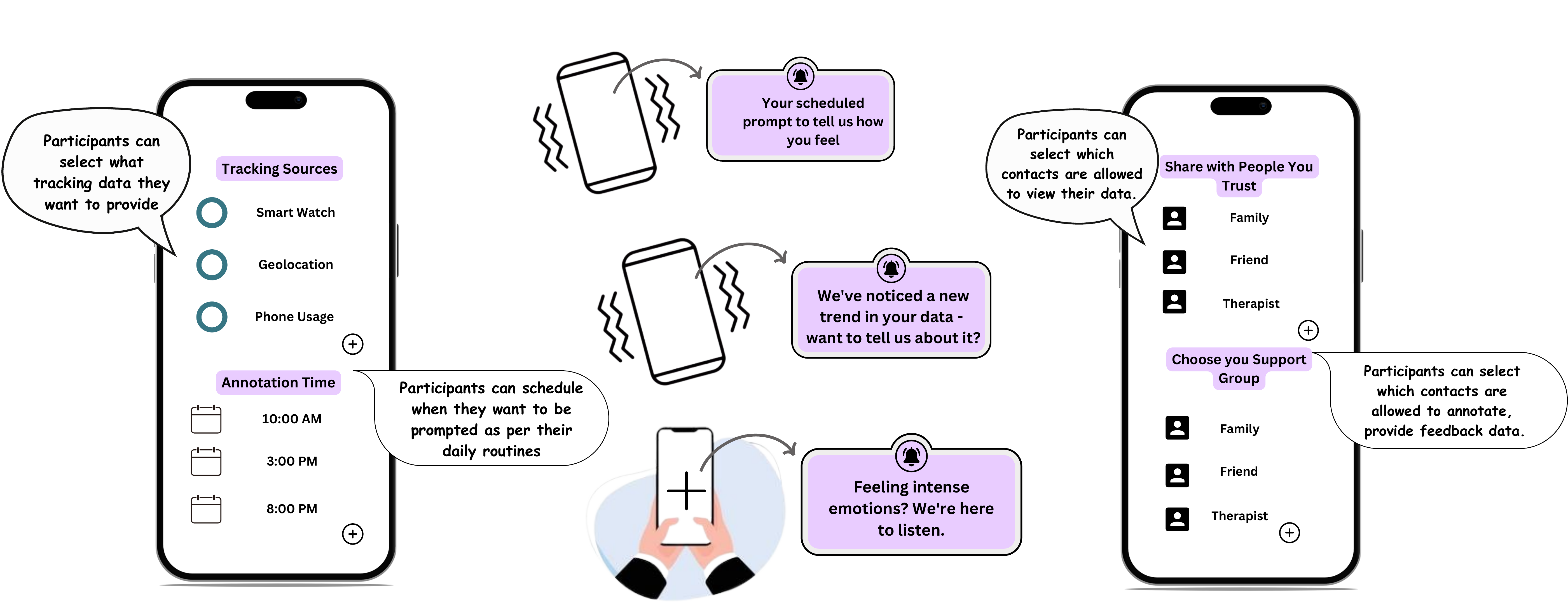} 
    \caption{Visualization of Integrating Participant Agency into the Design Process. The first screen illustrates how participants can exercise agency by selecting preferred data sources and specifying suitable time slots for receiving prompts based on their individual schedules. The second screen presents three prompting strategies: (1) prompts delivered at user-specified times, (2) context-aware prompts triggered by physiological or behavioral indicators, and (3) user-initiated annotations during emotionally salient moments. To further support multi-perspective reflection, participants are also given the option to include input from trusted members of their support network.}
    \label{fig:annotation_visual2}
\end{figure*}

\begin{figure*}[ht]
    \centering
    \includegraphics[width=0.8\textwidth]{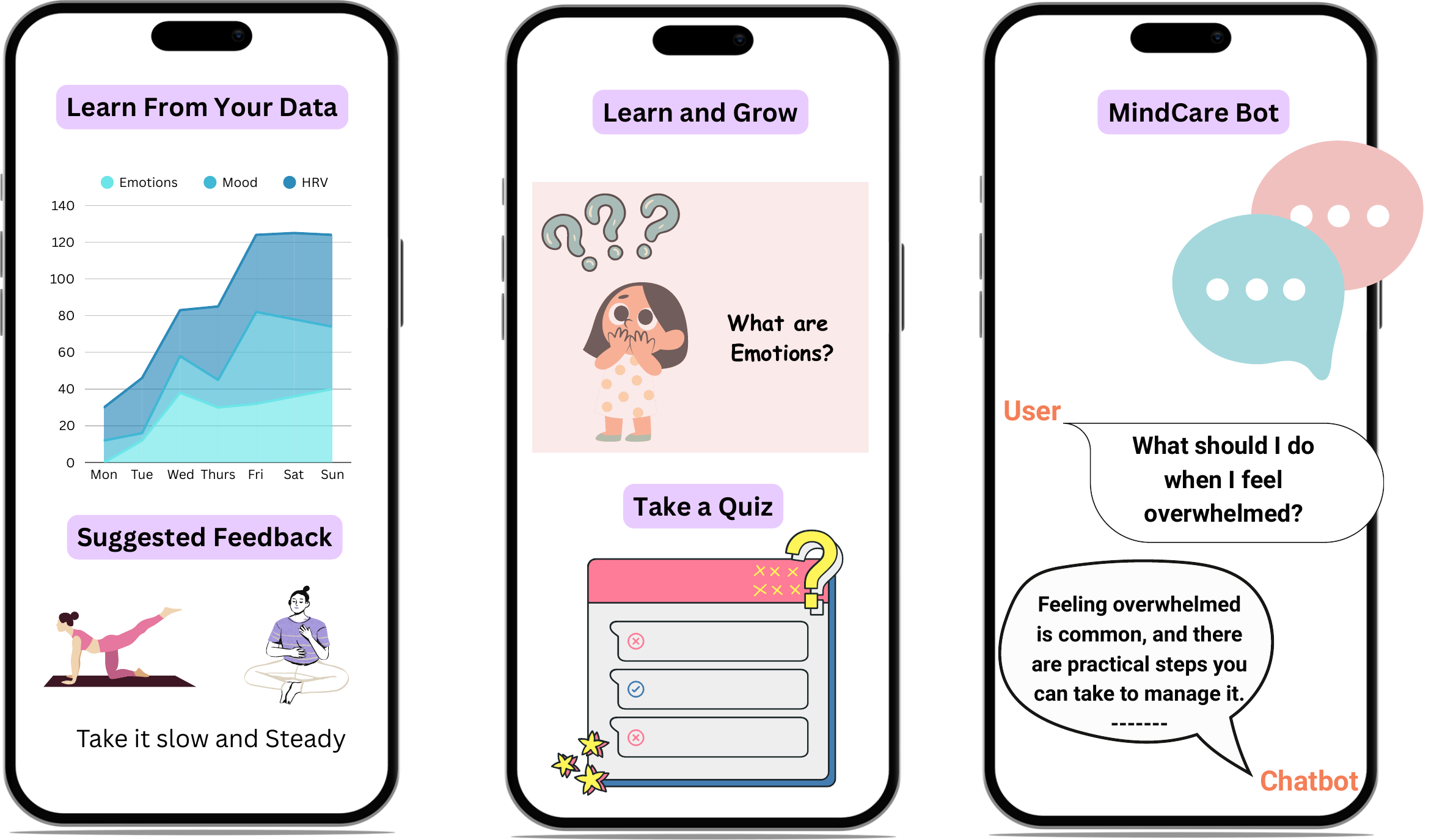} 
    \caption{Visualization of Participant Engagement, Learning, and Support Elements Integrated into the Design. The first screen displays personalized data insights derived from participants’ inputs to foster self-reflection. The second screen offers curated, trustworthy information aimed at enhancing emotional awareness and literacy. The third screen illustrates how LLM-supported systems can be incorporated to provide contextually relevant guidance and emotional support.}
    \label{fig:annotation_visual3}
\end{figure*}

\subsection{Implementing \textit{AnnoSense}: Designing for Participants}\label{prototype}

Building on the \textit{AnnoSense} framework, in this section, we envision potential directions for prototyping new tools that can further support both real-life emotion data collection and the advancement of wearable and mobile-based AI solutions. 
To inform the design of potential prototypes, we reviewed the designs of currently available mobile applications and wearable technologies that support mood tracking, mental health monitoring, and emotion-related interventions. This included mood and journaling apps such as MoodPrism \cite{rickard2016development}, MetricWire \cite{metricwire}, Daylio \cite{daylio2024}, and MindLamp 2 \cite{vaidyam2022enabling}; well-being and mindfulness platforms like Headspace \cite{headspace2024}, Happify \cite{happify, howells2016putting}, and Calm \cite{calm2024}; as well as wearable ecosystems including Apple Health \cite{applewatch2024}, Samsung Health \cite{samsungwatch2024}, Fitbit \cite{fitbit2024}, Oura Ring \cite{ouraring2024}, and WHOOP \cite{whoop2024}. We also examined AI-supported mental health applications such as Wysa \cite{wysa2024} and Woebot \cite{woebot2024}. Additionally, we also reviewed well-know EMA frameworks like MindLamp \cite{vaidyam2022enabling}, Beiwe \cite{onnela2021beiwe}, AWARE \cite{ferreira2015aware}, PACO \cite{paco}, Sensingkit \cite{katevas2016sensingkit}, mEMA \cite{ilumivu_mema}, Experiencesampler \cite{thai2018experiencesampler}, and MobileQ \cite{meers2020mobileq}. Our review identified several opportunities for designing future emotion annotation tools. 
Based on our review and the AnnoSense guidelines, we propose a set of prototype interfaces to accommodate users’ needs.

\textbf{1) Adaptable Annotation Interface:} We propose a prototype for adaptive annotation interfaces that provides users with an opportunity to select an annotation method according to their emotional intensity or available time, as illustrated in Figure \ref{fig:annotation_visual1}. This approach, in contrast to traditional ESM methods, provides users with an option to select between quick scale-based annotations, detailed subjective annotations \cite{10.1145/3699761, 10.1145/3613904.3642937,li2024diaryhelperexploringuseautomatic, kim2024healthllmlargelanguagemodels}, and chatbot-supported approaches \cite{10.1145/3613904.3642790, nepal2024contextual, 10.1145/3699761}. It also offers users appropriate guidance through curated emotion lists based on their selected overall feelings during quick annotations. Diary-inspired reflection prompts based on psychological frameworks, such as the ABC model (A - Activating Event, B - Beliefs, C- Consequence) \cite{malkinson2010cognitive} or Cognitive Behavioral Therapy (CBT) thought record model (Triggering event, Automatic thoughts, Emotions, Evidence supporting, and Evidence against) \cite{li2024diaryhelperexploringuseautomatic, nepal2024contextual, Kim_2024, wenzel2017basic}. And an empathetic chatbot interface to support emotion annotations.
Additionally, it supports a diverse range of users by offering options to record audio and upload images as part of their emotion annotations.
Furthermore, these interfaces can be designed with an added redundancy layer to enhance data collection consistency. This redundancy can be implemented by adding a quick-annotation option in each annotation mode, ensuring a baseline level of information, followed by options to add deeper reflections. This structure helps maintain a consistent data format while supporting varying user engagement levels. Moreover, the interface can also contain a curated list of activities that users can select to suggest the situational context of their data.

\textbf{2) User-Powered Interface for Prompting and Multi-Source Assessments:} We propose an emotion co-annotation platform where users can choose what data to share, set personalized prompting schedules, and invite trusted individuals to contribute their perspectives, as illustrated in Figure \ref{fig:annotation_visual2}. These tools can allow users to set personalized prompting conditions (e.g., time-based, data-triggered \cite{karapanos2012beyond, 10.1145/3699755, kaye2015dynamic, merrill2024transformingwearabledatahealth}, or self-initiated) and control the granularity of the emotion data they wish to share. Overall, such systems can enhance user agency. Moreover, an additional Likert scale to provide feedback on the confidence of emotional assessments can also be incorporated \cite{schroder2006first} for users and other sources. This additional confidence assessment from various sources can help recognize the validity of data and provide users with an additional layer of reflection on their annotations.

\textbf{3) Interface to Accommodate Learning and Support:} We propose adding data insights, verified sources for emotional well-being and awareness guidance, and LLM-supported guidance systems to the data-collection applications. These design elements can enhance user engagement by motivating them to understand themselves better, as illustrated in Figure \ref{fig:annotation_visual3}. 

Lastly, future work can explore a range of mood tracking and self-reflection applications that prioritize adaptability, user privacy, data sharing, and personalized emotional insight. For example, iMoodJournal \cite{imoodjournal_app} allows users to select their mood from an extensive list of emotions and supplement entries with journal notes, images, and location tags. It supports mood log sharing while maintaining a strong emphasis on data privacy. Similarly, Apple Health \cite{applewatch2024} offers a “State of Mind” mood logging feature, which prompts users to first categorize their mood as positive or negative, then select specific emotions with the option for multiple selections. Users can also add contextual notes and identify potential causes of mood changes, such as activities and relationships, while choosing between real-time and daily summary logging. Additional examples of feedback-oriented platforms include Mindsera \cite{mindsera_app}, an AI-powered journaling app that provides emotion analysis and personalized suggestions to guide self-reflection and promote mental fitness. Another example is Daylio \cite{daylio2024}, which offers a quick and streamlined interface for logging moods and activities multiple times throughout the day. Overall, platforms that integrate mood tracking with everyday lifestyle have the potential to generate more accurate, actionable emotion data and offer valuable insights for both research and personal well-being.

\subsection{Understanding the Implications of Pre-Study Guidelines}

\begin{quote}
    \textit{"I am made of little rooms full of thoughts, emotions \& memories. You cannot define me by listening to me once. I’m too complex." }
    [Source: Unknown, Credit: Pinterest]
\end{quote}

Prior research has often relied on one-way communication for everyday emotion data collection, borrowing heavily from traditional lab-based methodologies. However, everyday settings differ significantly from lab environments, as they lack the level of control typically available in laboratories. This lack of control introduces challenges in ensuring the quality and reliability of the collected data. To overcome these challenges, it is crucial to prepare a data collection pipeline that is robust to noise and bias in the real world. Beyond the lack of control, our findings also highlighted the diversity in participants' attitudes toward collecting and sharing emotional data, influenced by varying levels of emotional literacy. This diversity has been shown to impact emotion tracking among participants in previous studies \cite{10.1145/3025453.3025750,10.1145/3411764.3445771,10.1145/3313831.3376362, 10.1145/3476049, 10.1145/3491102.3517498}.
Drawing from data-centric AI guidelines \cite{10.1145/3571724}, ethical considerations for emotion AI \cite{10.1145/3442188.3445939}, and past literature on emotion tracking, our findings emphasize the need for careful pre-preparation which involves: 1) \textbf{careful selection of participants} (G1), by clearly defining the inclusion and exclusion criteria, selecting participants from diverse backgrounds \cite{10.1145/3569483}, and screening for possible conditions that could impact the data, 2) \textbf{preparing the participants} by elaborately informing (G2, G12), and training (G3–G4) them about how to interact efficiently with devices and annotation methods involved in data-collection methodology and its benefits, 3) \textbf{understanding the participants emotional profile} by performing elaborate psycho-social profiling (G5) and comprehensive demographic data collection (G6). Including these steps in the data collection pipeline can enhance participant engagement but also minimize errors, reduce hesitations, and foster a sense of collaboration between researchers and participants, which was often missing in prior methods \cite{kang2023k, 10.1038/s41597-021-00945-4, smets2018large, hosseini2022multimodal}. This careful pre-preparation will ensure an inclusive experience for participants of varying levels of emotional awareness. 
Furthermore, gathering psycho-social profiles and comprehensive demographic data will allow researchers to collect broader context about emotional responses missing in prior context collection that was limited to activity levels, basic demographics, and personality traits \cite{10.1109/TAFFC.2016.2625250, kang2023k} and will further help researchers to tailor the data collection process to the participants' emotional traits and lifestyles. 

\subsection{\textit{Go with the Flow}: Understanding the Implications of During Data-Collection Guidelines}

Prior research on emotion data collection has typically focused on two approaches for collecting emotion data in real-life settings. 1) Designing real-life emotional scenarios, such as work-related stress \cite{smets2018large, hosseini2022multimodal}, group entertainment \cite{bota2024real}, or driving stress \cite{healey2005detecting}. 2) Complete in-situ settings- where data collectors rely completely on participants' willingness to complete ecological momentary assessments (EMAs) or emotion questionnaires \cite{kang2023k, 10.1038/s41597-021-00945-4}.
These approaches often lead to data of a specific emotional scenario or incomplete data with limited contextual information. However, our findings emphasized the \textbf{inherent diversity in participants' attitudes} towards tracking and sharing emotion data and suggested designing sampling strategies that can accommodate this diversity. To address the challenges posed by the diverse engagement styles, we propose designing annotation methods that are tailored to the specific needs and capabilities of different individuals while also being flexible enough to support a broad range of participants. We recommend designing adaptable methods that can accommodate the varying needs of people, as recommended in section \ref{prototype}. Additionally, for participants with lower emotional literacy, researchers can frame emotional tools as practical aids rather than self-reflective interventions (e.g., stress relief or productivity enhancers) to increase engagement. Subsequently, for clinical participants or participants dealing with emotional trauma or other life-changing events, researchers can investigate a multiple-assessment approach \cite{10.1145/2968219.2968301}. Furthermore, our findings show that participants preferred using objective methods (such as scales) and lower frequencies on neutral days, while subjective methods (like written descriptions) and higher frequencies were favored during periods of intense emotions. However, present approaches such as ESM (In-the-moment annotation) and DRM (after-the-fact annotation) \cite{10.1145/3334480.3383019, 10.1016/j.ijhcs.2018.12.002, kosch2020emotions, schneider2020comparability, stone2006population} often overlook this fluidity in emotional experiences. These methods use either a fixed scale (e.g., SAM, Likert Scale) or questionnaires (e.g., STAI, PHQ-9) to capture emotion ratings within fixed or random time periods. This often doesn’t provide users with an opportunity to label as per emotion intensity, thus leading to datasets that fail to capture the dynamic and contextual aspects of emotional experiences, instead treating emotions as discrete snapshots, to overcome these challenges we recommend designing systems that can adapt to users changing emotional landscapes (G9), as shown in Figure \ref{fig:annotation_visual1}. Moreover, the timing of the annotation prompt can significantly affect the precision of annotations (G7, G8). For instance, in-the-moment annotation requires participants to assess and record their emotions as they occur, which can capture more immediate and authentic emotional states. However, this method can be cognitively demanding, as participants need to be aware of their emotions while balancing other activities in their environment \cite{10.1145/3334480.3383019, 10.1016/j.ijhcs.2018.12.002, kosch2020emotions}. In contrast, after-the-fact annotation allows participants to reflect on their emotional experiences once they have passed, which can provide a more thorough and considered response. However, this retrospective approach comes with its own cognitive challenges: memory bias and difficulty in recalling the intensity or nuances of past emotions accurately \cite{schneider2020comparability, stone2006population}. This can lead to data that may not fully reflect the emotional state experienced at the time, impacting the validity of the data for training AI systems. We recommend future works to design participant-aware sampling techniques and adaptable annotation methods that combine closed-end and open-ended questions (G9) as shown in Figures \ref{fig:annotation_visual1} and \ref{fig:annotation_visual2}. Further, an interface for adding contextual metadata, such as associated events or environmental factors, alongside emotional labels \cite{10.1145/3544549.3573803, smets2018large}, should also be added. Finally, our data also shed light on the \textbf{psychological influences that self-reporting emotions} can have on participants' daily lives. While self-reporting can foster self-awareness and provide emotional patterns, it can also influence the user’s emotions in unintended ways. For instance, users shared that recording subtle negative emotions can amplify overthinking. Conversely, documenting positive emotions can foster a sense of gratitude. To overcome potential influences, we recommend designing supportive and non-judgmental annotation techniques (G9, G11). For example, adding reflective prompts to encourage users to frame negative emotions constructively, like “What can this feeling teach me?”. Further, LLM-based structured journaling activities, guidance for mindfulness or relaxation exercises, and references to resources during distressing periods can be added to the applications \cite{10.1145/3699761, 10.1145/3491101.3519854}. Further tools can integrate features that allow users to record emotions without immediate analysis and then review entries after a period of detachment, or ask participants to note a small positive event or something they feel grateful for (G11), can be added, as shown in the prototype figure \ref{fig:annotation_visual3}.

\subsection{Moving beyond the Traditional Data Modeling: Implications on Post-Data Collection Approaches}

Our findings emphasize that the concept of \textbf{"emotional ground truth"} extends far beyond the survey responses typically gathered through standard questionnaires. However, current datasets often assume a universal definition of emotions and one-to-one mappings between emotion data and filled surveys \cite{barrett2017emotions, singh2024saycatcatunderstanding, gao2023critiquing}, to label emotion data. Moreover, prior work on developing models for physiological emotion data often applies simplistic labeling approaches like categorizing emotions into discrete groups based on objective labels, such as predefined emotion categories (e.g., happy, angry) or scales (e.g., 1 to 5). These methods often do not use additional contextual data \cite{10.1145/3242969.3242985, 10.1109/TAFFC.2016.2625250, 10.1038/s41597-019-0209-0, 9779458}, while modeling the AI algorithms leading to the development of models that cannot be adapted in real-life \cite{hickey2021smart} or clinical settings \cite{abd2023wearable, abd2023systematic}. The continued use of such approaches can be attributed to several factors: 1) Simplifying emotion categorization reduces the complexity of emotion recognition models, making them easier to develop, train, and implement.
2) Discrete emotion categories are easier for participants or experts to label, lowering the annotation burden.
3) Standardizing datasets based on these categories facilitates generalization across various AI applications, such as sentiment analysis and video emotion recognition.
4) The influence of early psychological theories, such as basic emotion theory, has strongly shaped these practices.
However, our findings based on interviews with domain experts challenge these assumptions. Experts argue that actionable insights and contextually relevant data should take precedence over overly generic labeling (G13, G14). They suggested that emotional ground truth is not a simple, one-to-one mapping from data to labels. In fact, it’s a composite representation that varies according to the user profile. 

Insights from both participants and experts have shaped our guidelines for collecting emotion data that is dynamic, layered, and actionable. Unlike traditional methods, our approach captures emotions in real-time and across varying contexts, resulting in data that is fundamentally different in structure and complexity. This shift highlights the need for new labeling and validation techniques that can accommodate the richness and variability of the collected data. Traditional emotion datasets often collect inputs, such as single-point self-reports, task-based annotations, or expert labels, resulting in relatively uniform data structures that are easy to label. These inputs are then reduced to binary or discrete categories (e.g., “happy” or “stressed”) by binning self-reported or task-driven
labels to fit downstream tasks. For example, in the WESAD dataset \cite{10.1145/3242969.3242985}, emotional states are classified into stress versus no-stress categories based solely on experimental stimuli, without incorporating participant self-reports. Similarly, GLOBEM \cite{xu_globem_2022} focuses
on depression detection as a downstream task, and ASCERTAIN \cite{10.1109/TAFFC.2016.2625250} performs arousal-valence classification based on self-reports. In contrast, our approach captures emotion as a dynamic, evolving state, influenced by contextual, physiological, and subjective factors as discussed in section \ref{findings}. This results in data that is more variable and multidimensional. For instance, each emotional annotation in our system may include a combination of quick scale ratings, emotion labels, option text/audio/image data, confidence scores, and contextual metadata. The structure and depth of these annotations can vary based on the user’s engagement and the intensity of the emotional experience. Such variability introduces both opportunities and challenges: while the data offers a more accurate and holistic view of emotional states, it also complicates traditional labeling and validation pipelines, which typically assume uniform input formats.

To effectively handle our dynamic data, we propose a set of new validation schemes that go beyond traditional practices. \textbf{1) Triangulated Validation}: In this technique, we can assign a final emotion score or label by combining information from multiple sources, such as scale-based self-reports, emotion-list, physiological signals, AI-generated or text annotations, images/audio annotations (optional), and contextual, peer, or expert feedback. Each of these sources can first be evaluated for coherence and reliability in a given context, and a confidence score can be assigned to them. For example, if a user provides a confident self-report, it might carry a higher weight of 0.9, while physiological signals with strong indicators could be weighted at 0.8, and AI-based reflections with uncertain text data might be assigned a lower weight, such as 0.5. All emotion representations are then aligned into a common format, such as a valence-arousal score or a set of discrete emotion categories. Finally, the final label can be computed using a weighted aggregation, such as a weighted average for numerical scores or a confidence-weighted majority vote for categorical labels. This ensures that more reliable sources contribute more to the outcome. Overall, this approach allows for a more robust and context-aware emotional label, addressing the limitations of relying on any single data source. \textbf{2) Semantic Validation}: Given the redundant nature of information collected through multiple methods, such as scale-based self-reports, emotion names from the list, or optional text, should be validated for semantics. This means making sure that elements like emotion labels, confidence scores, and multimedia content (such as text, images, or audio) align coherently. For example, if an annotation includes the emotion label “joy,” but the accompanying text expresses sadness or the image shows someone crying, a mismatch may need to be addressed. Similarly, if users rate their emotional intensity as very high but give a very low confidence score, that inconsistency could indicate confusion or noise in the data. This form of validation can add a layer of reliability in collected data, which is often missing in traditional data. \textbf{3) Contextual validation}: This involves checking whether the data fits logically within the context in which it was collected. This validation technique is similar to traditional approaches of checking the data contextually. \textbf{4) Annotation Agreement Validation and Co-Development:} This validation focuses on assessing the consistency of emotion annotation when multiple annotators (participants, experts, and peer groups) are involved in labeling the same content. Since emotions are highly personal and subjective, it's common for different users to interpret the same situation differently. This step helps identify how much agreement or disagreement exists among annotators. Techniques like inter-annotator agreement metrics (e.g., Cohen's Kappa or Krippendorff's Alpha) can be used by future works to quantify the level of consistency across annotations. This approach also provides a framework for emotion data collectors, mental health professionals, and emotion AI experts to co-develop and evaluate new tools and methodologies with users. By bringing together multiple stakeholders in the data validation process, the resulting systems can benefit from diverse expertise: users' lived experiences, clinicians' domain knowledge, and technical experts' implementation capabilities. This collaborative approach ensures data collection tools are not only technically sound but also clinically relevant and ethically implemented. Consequently, these validation techniques can validate emotion data more robustly, and they also align with domain experts' guided strategies of finding \textit{congruence} in emotional assessments. Further, they provide a platform to add clinical and participant insights to traditional data, thus adding an opportunity for co-development with experts while keeping participants in the loop.
Lastly, our findings underscore the critical need to design AI algorithms that prioritize actionable outcomes \cite{10.1145/3699755}, such as identifying meaningful patterns—like recurring emotional states—over simplistic emotional categorizations \cite{10.1145/3544548.3581209, 10.1145/3491102.3517498}. This shift is essential for developing systems that align more closely with real-world applications. For example, in therapeutic settings, recognizing patterns in emotional states over time can help identify triggers or trends in mental health, providing valuable insights for personalized interventions. However, for clinical settings, tracking changes in symptoms can be targeted over nuanced emotional changes. By focusing on actionable outcomes, algorithms can also provide deeper insights for decision-making, enabling stakeholders to address the underlying causes of emotional responses rather than just classifying emotions into predefined categories. This approach moves away from a rigid framework of labeling emotions, embracing a more dynamic and context-sensitive model of emotion tracking.

\section{Limitations}
This study aimed to explore people’s attitudes and preferences toward tracking and monitoring emotions in everyday settings for emotion AI data collection and interventions, and provide a set of guidelines for future emotion data collection methods. A limitation of our work was that most of our participants were well-educated, tech-savvy individuals familiar with AI, emotion monitoring, and wearable technologies. Thus, our findings might not generalize well to people with low literacy and less experience with emotion tracking. We also recognize that our findings might be influenced by the participants' demographics, group composition, and backgrounds, since all our participants belonged to the same cultural background and country. To address this, we have included a diverse audience of users and non-users of emotion-tracking technology with varying levels of technological familiarity, emotional awareness, and demographic profiles (age, gender, occupation, education). 
Moreover, it is also important to recognize that different research objectives may encounter unique challenges when adapting these guidelines. Thus, we recommend that future work customize these guidelines to their specific needs for better adaptability. For instance, participant training and psycho-social profiling can be significantly more challenging when working with clinical populations compared to undiagnosed, healthy counterparts. Individuals with diagnosed mental disorders may require tailored approaches to ensure ethical considerations, comfort, and engagement throughout the data collection process. To address these complexities, we recommend engagement with mental health professionals to navigate the sensitivities associated with clinical populations.  This is particularly crucial given that many emotion monitoring interventions are designed to target individuals with diagnosed mental health conditions. By including professional supervision, researchers can better align their methodologies with the needs of clinical audiences, creating a more inclusive, ethical, and effective data collection process \cite{10.1145/3699755} tailored to diverse participant groups.

\section{Conclusion}

Our study investigated the perspectives of key stakeholders (public and mental health professionals) on annotating emotion data as part of everyday life. Previously, emotion data collection relied on approaches based on objective scales and questionnaires, both in laboratory settings and real-world environments. However, the impact of user-specific factors and the influence of everyday contexts on the quality of emotion annotations has been largely understudied. Our analysis reveals that factors such as the fluidity of emotional experiences, stigma, and emotional literacy significantly affect the accuracy of these annotations. By examining these factors, our study provides a comprehensive understanding of their implications on data collection in everyday settings. Based on these insights, we offer a framework \textit{AnnoSense} for future works to develop more holistic approaches for emotion data collection. In the future, we aim to expand these design guidelines into practical solutions and algorithms suitable for daily life settings, exploring their effectiveness in future solutions for wearable and mobile-based emotion AI systems.

\section{Acknowledgements}

We gratefully acknowledge Dr. Koushik Sinha Deb from the Department of Psychiatry at AIIMS New Delhi for his valuable support in recruiting focus group participants. We also extend our sincere thanks to all participants and evaluators for their time, openness, and for sharing their insights throughout the study. And acknowledge the support of the iHub-Anubhuti-IIITD Foundation, established under the NM-ICPS scheme of the DST at IIIT-Delhi.

\bibliographystyle{ACM-Reference-Format}
\bibliography{sample-base}

\pagebreak

\begin{appendices}

\section{Semi-Structured Interview Guide}\label{interviewapp}

This appendix presents the semi-structured interview guide used to explore participants’ experiences, perceptions, and preferences related to emotion annotation. Given the semi-structured nature of the interviews, the questions were adapted as needed to ensure clarity and comprehensibility for participants. The guide is organized according to the \textit{Who, What, When, Where, Why,} and \textit{How} framework, followed by additional probing questions.

\smallskip

\noindent\doublebox{
\begin{minipage}{0.95\textwidth}
\textbf{WHO: Participant Background and Emotional Self-Reflection}

\vspace{0.5em}
\textbf{Self-Reflection on Emotions}
\begin{itemize}[leftmargin=*]
    \item Could you tell me a bit about yourself, particularly in terms of how you experience and relate to emotions?
    \item How would you describe your emotional landscape and the role emotions play in your daily life?
    \item How do you perceive your ability to manage or process emotions?
    \item Would you describe yourself as more emotionally expressive or emotionally reserved?
    \item How do you typically respond to emotional experiences?
    \item To what extent would you consider yourself emotionally self-aware?
\end{itemize}

\vspace{0.5em}
\textbf{Familiarity with Technology}
\begin{itemize}[leftmargin=*]
    \item How familiar are you with using digital technologies, such as mobile applications or wearable devices, for tracking or annotating emotions?
\end{itemize}

\vspace{0.5em}
\textbf{Experience with Emotion Annotation}
\begin{itemize}[leftmargin=*]
    \item Have you had any prior experience with tracking or annotating your emotions?
    \item Have you used any specific tools or methods—such as journaling, mood-tracking apps (e.g., Likert scales, emojis), or verbal/voice recordings—to annotate emotions?
\end{itemize}

\vspace{0.5em}
\textbf{Psychological Impact of Annotation}
\begin{itemize}[leftmargin=*]
    \item How does the process of annotating emotions affect you psychologically?
    \item Do you find it therapeutic, stressful, or something else?
    \item In what ways does it influence your emotional awareness and understanding?
\end{itemize}

\vspace{0.5em}
\textit{Note to participants: ``Emotion annotation refers to the practice of labeling or recording emotional states, often to support self-reflection, research, or the training of AI systems.''}
\end{minipage}
}

\noindent\doublebox{
  \parbox{0.95\linewidth}{
    \textbf{WHAT: Content and Scope of Annotation}

    \vspace{0.5em}
    \begin{itemize}[leftmargin=*]
        \item What kinds of emotions do you think should be annotated?
        \item Which emotional states or types of experiences do you believe are most important to capture?
        \item Can you provide examples of specific situations or emotional experiences that you would consider annotating?
        \item Do you have any privacy concerns regarding emotion annotation? Would you feel comfortable annotating deeply personal emotions in detail?
    \end{itemize}
  }
}

\noindent\doublebox{
  \parbox{0.95\linewidth}{
   \textbf{WHEN: Timing of Emotion Annotation}

   \vspace{0.5em}
    \begin{itemize}[leftmargin=*]
    \item When do you think is the most appropriate time to annotate emotions?
    \item (For participants with prior experience) When do you typically annotate your emotions, and in what kinds of scenarios?
    \item Would you prefer to annotate emotions in real-time (immediately after experiencing them), or retrospectively (e.g., summarizing emotions at the end of the day)? Why?
    \item How frequently do you believe emotional annotation should occur?
    \end{itemize}
 }
}

\noindent\doublebox{
  \parbox{0.95\linewidth}{
   \textbf{WHERE: Context and Environment for Annotation}

   \vspace{0.5em}
    \begin{itemize}[leftmargin=*]
        \item In what types of environments would you feel most comfortable annotating your emotions?
        \item Would you prefer to annotate emotions at home, in the workplace, or in another setting? Why?
        \item Are there any places or contexts where you would feel uncomfortable annotating emotions?
        \begin{itemize}
            \item If participant responds ``alone,'' follow up with: ``If you are unable to be alone—e.g., at work or in a public space—would you still feel comfortable annotating?''
        \end{itemize}
        \item Can you describe a scenario in which annotating emotions would be particularly difficult?
        \begin{itemize}
            \item What factors would contribute to that difficulty?
            \item How might you address or overcome them?
        \end{itemize}
    \end{itemize}
 }
}

\noindent\doublebox{
  \parbox{0.95\linewidth}{
   \textbf{WHY: Motivation and Perceived Value}

   \vspace{0.5em}
    \begin{itemize}[leftmargin=*]
        \item Why do you think annotating emotions is important or meaningful?
        \item What personal benefits do you associate with the annotation of positive or negative emotions?
        \item What challenges or barriers do you foresee in the emotion annotation process?
    \end{itemize}
 }
}

\noindent\doublebox{
  \parbox{0.95\linewidth}{
   \textbf{HOW: Preferred Methods and Tools for Annotation}

   \vspace{0.5em}
    \begin{itemize}[leftmargin=*]
        \item How would you go about annotating your emotions?
        \item What tools or methods would you prefer to use (e.g., paper journals, apps with Likert scales or emojis, voice recordings)?
        \item How much time would you be willing to dedicate to emotion annotation per day or week?
        \item What features or types of support would make the annotation process easier or more engaging?
        \item Are there specific functions or aids (e.g., reminders, visualizations, AI feedback) that would help you annotate more effectively?
        \item How could the emotion annotation process be simplified or made more intuitive?
    \end{itemize}
 }
}

\noindent\doublebox{
  \parbox{0.95\linewidth}{
   \textbf{Additional Questions}

   \vspace{0.5em}
    \begin{itemize}
    \item What are your overall expectations from the annotation process?
    \begin{itemize}
        \item What outcomes do you hope to achieve?
        \item How would you evaluate or measure the success of the process?
    \end{itemize}
\end{itemize}
 }
}

\section{Focus Group Discussion}\label{FGDAPP}

\noindent\doublebox{
  \parbox{0.95\linewidth}{

    \vspace{0.8em}
    \textbf{Introduction}
    \begin{itemize}[leftmargin=*]
        \item Brief overview of the study and purpose with presentation.
        \item Warm-up conversation and informed consent.
    \end{itemize}

    \vspace{0.8em}
    \textbf{Current Practices for Assessing Emotional States}
    \begin{itemize}[leftmargin=*]
        \item How do you currently assess the emotional states of your patients?
        \item What tools or techniques do you use to collect data on your patients' emotions?
    \end{itemize}

    \vspace{0.8em}
    \textbf{Attitude towards Data and AI}
    \begin{itemize}[leftmargin=*]
        \item Can you elaborate on what AI tools you would like to use in your practice?
        \item What are your initial thoughts on the use of AI to understand and monitor emotions?
        \item How do you think AI can enhance emotional well-being and mental health care?
        \item What are the potential benefits and drawbacks of using AI for emotional recognition in clinical settings?
    \end{itemize}

    \vspace{0.8em}
    \textbf{Emotion Data Collection}
    \begin{itemize}[leftmargin=*]
        \item What are opportunities for the present ways of emotion annotations (as presented in the introduction), and why, according to you?
        \item Which emotions do you believe are important to track daily to maintain good mental well-being?
        \item In your experience, how easy is it for individuals to understand their emotions?
        \item Do you think the process would be more challenging for people who are emotionally susceptible or are suffering from some minor disorders?
        \item What do you see as the main challenges in collecting emotion data in everyday settings?
        \item What should we call the "emotion ground truth" or "emotion label," and why?
        \item At what resolution (e.g., frequency, granularity) should we track emotions to make effective interventions?
    \end{itemize}
  }
}

\section{Survey Questionnaire}\label{surveyapp}

\begin{center}
\setlength{\fboxsep}{10pt}
\doublebox{
\begin{minipage}{0.95\textwidth}

\vspace{0.8em}

\textbf{Section 1: Participant Background}
\begin{itemize}
    \item Consent to participate
    \item Age, Gender, Education, Occupation
\end{itemize}

\vspace{0.5em}

\textbf{Section 2: Understanding Emotional Awareness}
\begin{itemize}
    \item \textbf{Q1.} How often do you reflect on your emotions?
    \item \textbf{Q2.} How easily can you identify emotions during strong experiences?
    \item \textbf{Q3.} How often do you feel mixed emotions?
    \item \textbf{Q4.} Do you use any tools (e.g., journaling, mood tracking apps)?
    \item \textbf{Q5.} Think about a recent time when you felt a strong emotion. What emotion did you feel? (Please write a brief description of the situation and the emotion you identified)
    \item \textbf{Q6.} Looking back at the situation you described in question above and how accurate do you think your emotion label was?
    \item \textbf{Q7.} Name up to 5 positive emotions you feel daily and their impact in your daily life.
    \item \textbf{Q8.} Name up to 5 negative emotions you feel daily and their impact and in your daily life.
    \item \textbf{Q9.} Which emotions are easiest to identify, and why?
    \item \textbf{Q10.} Which emotions are hardest to identify, and why?
    \item \textbf{Q11.} Can you differentiate between similar emotions (e.g., sadness vs. disappointment or anger and frustration)? Explain how?
    \item \textbf{Q12.} What does the intensity of an emotion mean to you? Explain?
    \item \textbf{Q13.} What best describes an "emotion" (select all that apply)?
\end{itemize}

\vspace{0.5em}

\textbf{Section 3: Attitudes Toward Daily Emotion Annotation}
\begin{itemize}
    \item \textbf{Q14.} How confident are you in labeling your emotions accurately?
    \item \textbf{Q15.} How well can you label mixed emotions?
    \item \textbf{Q16.} How would you prefer to annotate emotions (e.g., text, emojis, scale)?
    \item \textbf{Q17.} Explain why you chose a particular option in Q16?
    \item \textbf{Q18.} What factors are most important when labeling emotions? (e.g., context, physical response)
    \item \textbf{Q19.} Why did you choose your preferred annotation method?
    \item \textbf{Q20.} Do cultural or societal factors influence your emotion labeling? Please explain.
    \item \textbf{Q21.} Would you like to annotate emotions daily?
    \item \textbf{Q22.} If Yes, Why would you like to annotate your emotions daily?
    \item \textbf{Q23.} If No, Why not would you like to annotate your emotions daily?
    \item \textbf{Q24.} How easy is daily emotion annotation for you?
    \item \textbf{Q25.} How frequently can you annotate emotions?
    \item \textbf{Q26.} Would you annotate negative emotions (e.g., anger, stress)? Why or why not?
    \item \textbf{Q27.} Would you annotate positive emotions (e.g., calm, joy)? Why or why not?
\end{itemize}
\end{minipage}
}
\end{center}

\begin{table}[!htbp]
\centering
\begin{tabular}{
    @{} 
    >{\centering\arraybackslash}p{3cm}
    @{\hspace{1pt}} 
    >{\centering\arraybackslash}p{2.2cm}
    @{\hspace{1pt}} 
    >{\centering\arraybackslash}p{2.2cm}
    @{\hspace{1pt}} 
    >{\centering\arraybackslash}p{2.2cm}
    @{\hspace{1pt}} 
    >{\centering\arraybackslash}p{6.8cm}
    @{} 
}
\toprule
\textbf{Question No.} & \textbf{Completed Response} & \textbf{Response Rate (\%)} & \textbf{Required Question} & \textbf{Word Count Summary} \\
\midrule
Q4 & 41 & 54.67 & Yes & Range = 1\textendash68,\quad \text{Mean} = 6.74,\quad \text{SD} = 15.23 \\
Q5 & 70 & 93.33 & Yes & Range = 1\textendash105,\quad \text{Mean} = 22.59,\quad \text{SD} = 24.44 \\
Q7 & 69 & 92.00 & Yes & Range = 1\textendash121, \quad \text{Mean} = 14.93,\quad \text{SD} = 26.28 \\
Q8 & 68 & 90.67 & Yes & Range = 1\textendash78, \quad \text{Mean} = 13.00,\quad \text{SD} = 17.49 \\
Q9 & 72 & 96.00 & Yes & Range = 1\textendash47, \quad \text{Mean} = 10.88,\quad \text{SD} = 10.91 \\
Q10 & 61 & 81.33 & Yes & Range = 1\textendash43, \quad \text{Mean} = 10.13,\quad \text{SD} = 10.58 \\
Q11 & 73 & 97.33 & Yes & Range = 1\textendash110, \quad \text{Mean} = 18.39,\quad \text{SD} = 21.32 \\
Q12 & 66 & 88.00 & Yes & Range = 2\textendash100, \quad \text{Mean} = 19.91,\quad \text{SD} = 17.37 \\
Q17 & 62 & 82.67 & No & Range = 2\textendash119, \quad \text{Mean} = 21.37,\quad \text{SD} = 20.20 \\
Q20 & 26 of 44 & 59.09 & No & Range = 5\textendash196, \quad \text{Mean} = 36.15,\quad \text{SD} = 40.51 \\
Q22 & 25 of 28 & 89.29 & No & Range = 3\textendash39,\quad \text{Mean} = 14.20,\quad \text{SD} = 10.19 \\
Q23 & 45 of 47 & 95.74 & No & Range = 1\textendash48, \quad \text{Mean} = 14.53,\quad \text{SD} = 11.66 \\
Q26 & 66 & 88.00 & No & Range = 1\textendash57, \quad \text{Mean} = 14.23,\quad \text{SD} = 11.96 \\
Q27 & 62 & 82.67 & No & Range = 1\textendash61, \quad \text{Mean} = 12.79,\quad \text{SD} = 10.97 \\
\bottomrule
\end{tabular}
\caption{Quantitative Summary of Open-Ended Responses in our Survey}
\label{tab:survey-quality}
\end{table}

\end{appendices}

\end{document}